%% file: main.tex
\documentclass[conference]{IEEEtran}
\IEEEoverridecommandlockouts
\usepackage{cite}
\usepackage{amsmath,amssymb,amsfonts}
\usepackage{algorithmic}
\usepackage{graphicx}
\usepackage{textcomp}
\usepackage{xcolor}
\def\BibTeX{{\rm B\kern-.05em{\sc i\kern-.025em b}\kern-.08em
    T\kern-.1667em\lower.7ex\hbox{E}\kern-.125emX}}

\usepackage{bm}
\usepackage{graphicx}
\usepackage[ruled,vlined,linesnumbered,commentsnumbered]{algorithm2e}
\usepackage{enumitem}
\usepackage{multirow}
\usepackage{tikz} 
\usepackage{cancel}
\usepackage{balance}
\usetikzlibrary{fit}
\usepackage{url}
\usepackage{hyperref}
\usetikzlibrary{positioning}
\usepackage{array}
\usetikzlibrary{external}
\newtheorem{theorem}{Theorem}
\newtheorem{lemma}{Lemma}
\newtheorem{proposition}{Proposition}
\newtheorem{definition}{Definition}
\newtheorem{example}{Example}
\newtheorem{problem}{Problem}[section]
\newcommand{\Lab}{\mathbb{L}}
\newcommand{\myIndex}{\texttt{RLC}}
\newcommand{\myQuery}{\texttt{RLC}}
\newcommand{\lin}{\mathcal{L}_{in}}
\newcommand{\lout}{\mathcal{L}_{out}}
    
\begin{document}

\title{A Reachability Index for Recursive Label-Concatenated Graph Queries}

\author{
\IEEEauthorblockN{
Chao Zhang\IEEEauthorrefmark{1}, Angela Bonifati\IEEEauthorrefmark{1},
Hugo Kapp\IEEEauthorrefmark{2}, Vlad Ioan Haprian\IEEEauthorrefmark{2} and
Jean-Pierre Lozi\IEEEauthorrefmark{2}}
\IEEEauthorblockA{\IEEEauthorrefmark{1}Lyon 1 University, Lyon, France}
\IEEEauthorblockA{\IEEEauthorrefmark{2}Oracle Labs, Zürich, Switzerland}
\IEEEauthorblockA{\{chao.zhang, angela.bonifati\}@univ-lyon1.fr, \{hugo.kapp, vlad.haprian, jean-pierre.lozi\}@oracle.com}
}

\maketitle
\thispagestyle{plain}
\pagestyle{plain}

\begin{abstract}
Reachability queries checking the existence of a path from a source node to a target node are fundamental operators for querying and processing graph data. Current approaches for index-based evaluation of reachability queries either focus on plain reachability or constraint-based reachability with only alternation of labels. 
In this paper, for the first time we study the problem of index-based processing for recursive label-concatenated reachability queries, referred to as \myQuery\ queries. 
These queries check the existence of a path that can satisfy the constraint defined by a concatenation of at most $k$ edge labels under the Kleene plus. 
Many practical graph database and network analysis applications exhibit \myQuery\ queries. However, their evaluation remains prohibitive in current graph database engines.

We introduce the \myIndex\ index, the first reachability index to efficiently process \myQuery\ queries.
The \myIndex\ index checks whether the source vertex can reach an intermediate vertex that can also reach the target vertex under a recursive label-concatenated constraint.
We propose an indexing algorithm to build the \myIndex\ index, which guarantees the soundness and the completeness of query execution and avoids recording redundant index entries. 
Comprehensive experiments on real-world graphs show that the \myIndex\ index can significantly reduce both the offline processing cost and the memory overhead of transitive closure, while improving query processing up to six orders of magnitude over online traversals.
Finally, our open-source implementation of the \myIndex\ index significantly outperforms current mainstream graph engines for evaluating \myQuery\ queries. 
\end{abstract}


\input{sections/Introduction}

\input{sections/Related_Works}
\input{sections/Preliminary}

\input{sections/Concatenation}
\input{sections/MyIndex}
\input{sections/Experimental_Results}
\input{sections/Conclusion}

\bibliographystyle{IEEEtran}
\bibliography{IEEEabrv,references}
\balance

\input{sections/Appendix}

\end{document}

%% file: sections/Introduction.tex
\section{INTRODUCTION}
Graphs have been the natural choice of data representation in various domains \cite{Newman2010}, \textit{e.g.}, social, biochemical, fraud detection and transportation networks, and reachability queries are fundamental graph operators \cite{10.1145/3186728.3164139}.
Plain reachability queries check whether there exists a path from a source vertex to a target vertex, for which various indexing techniques have been proposed \cite{10.1145/67544.66950,10.1145/99935.99944,10.5555/545381.545503,1617443,4497498,10.1145/1376616.1376677,10.1145/1559845.1559930,10.1145/2463676.2465286,10.14778/2556549.2556578,10.5555/1083592.1083651,10.1145/1247480.1247573,10.14778/1920841.1920879,6544893,Veloso2014ReachabilityQI,10.14778/2732977.2732992,7750623, 10.1145/2588555.2612181, 10.1145/1871437.1871457,10.1145/2505515.2505724}.
To facilitate the representation of different types of relationships in real-world applications, \textit{edge-labeled graphs} and \textit{property graphs}, where labels can be assigned to edges, are more widely adopted nowadays than unlabeled graphs.
Such advanced graph models allow users to add path constraints when defining reachability queries, which play a key role in graph analytics. 
However, current index-based approaches focus on constraint-based reachability with only alternation \cite{10.1145/1807167.1807183,ZOU201447,10.1145/3035918.3035955,10.14778/3380750.3380753,10.1145/3451159}. 
In this paper, we consider for the first time reachability queries with a complex path constraint corresponding to a \textit{concatenation} of edge labels under the Kleene plus, referred to as \textit{recursive label-concatenated queries} (\myQuery\  queries). 
\myQuery\ queries are a prominent subset of regular path queries, for which indexing techniques have not been understood yet. As such, they are significant counterparts of queries with alternation of labels as path constraints \cite{10.1145/3035918.3035955} and queries without path constraints, \textit{i.e.}, plain reachability queries \cite{10.1145/67544.66950}.
We further motivate \myQuery\ queries by means of a running example.

\begin{figure}
	\centering
\begin{tikzpicture}[node distance=25mm, thick, node/.style = {draw, circle,inner sep=2,outer sep=2, scale = 0.75}, label/.style = {scale = 0.7, sloped, midway}] 
	\node[node] (10)  {$P_{10}$};
	\node[node] (11) [right of=10] {$P_{11}$};
	\node[node] (12) [right of=11] {$P_{12}$};
	\node[node] (13) [right of=12] {$P_{13}$};
	\node[node] (14) [below of=10] {$A_{14}$};
	\node[node]  (15) [below of=11] {$E_{15}$};
	\node[node] (17) [below of=12] {$A_{17}$};
	\node[node] (16) [right of=13] {$P_{16}$};
	\node[node] (18) [right of=17] {$E_{18}$};
	\node[node] (19) [right of=18] {$A_{19}$};
	\draw[->] (10) -- node[label][above] {\texttt{knows}}(11); 
	\draw[->] (11) -- node[label][above] {\texttt{worksFor}}(12); 
	\draw[->] (11) to [out=45,in=180-45,looseness=0.7]node[label][above] {\texttt{knows}}(12); 
	\draw[->] (12) to [out=30,in=180-30,looseness=0.7]node[label][above] {\texttt{knows}}(16); 
	\draw[->] (13) to [out=180+20,in=360-20,looseness=1]node[label][below] {\texttt{knows}}(11);
	\draw[->] (12) -- node[label][above] {\texttt{knows}}(13); 
	\draw[->] (10) -- node[label][above] {\texttt{holds}}(14); 
	\draw[->] (14) -- node[label][above] {\texttt{debits}}(15);
	\draw[->] (13) -- node[label][above] {\texttt{worksFor}}(16);  
	\draw[->] (13) to [out=360-45,in=180+45,looseness=0.7]node[label][above] {\texttt{knows}}(16);
	\draw[->] (15) -- node[label][above] {\texttt{credits}}(17); 
	\draw[->] (16) -- node[label][above] {\texttt{holds}}(19); 
	\draw[->] (17) -- node[label][above] {\texttt{debits}}(18);
	\draw[->] (18) -- node[label][above] {\texttt{credits}}(19); 
\end{tikzpicture}
	\caption{A social and professional network for illustrating \myQuery\ queries.}
	\label{fig: example in introduction}
	\vspace*{-0.5cm}
\end{figure}
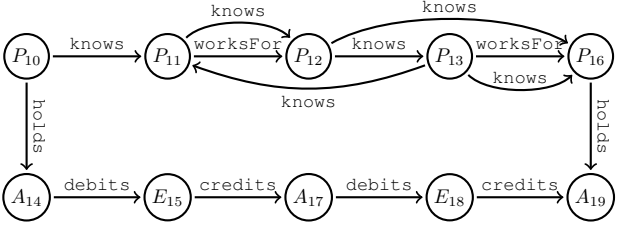

\begin{example}
Figure \ref{fig: example in introduction} shows a property graph inspired by a real-world use case encoding an interleaved social and professional network along with information of bank accounts of persons.
\myQuery\ queries can be used to detect fraud and money laundering patterns among financial transactions.
For instance, the query $Q1(A_{14},A_{19},($\texttt{debits, credits}$)^+)$ checks whether there is a path from account $A_{14}$ to $A_{19}$ such that the label sequence of the path is a concatenation of an arbitrary number of occurrences of $(\texttt{debits, credits})$, which can lead to detect suspicious patterns of money transfers between these accounts.
The \myQuery\ query  $Q1((A_{14}, A_{19}, ($\texttt{debits, credits}$)^+)$ evaluates to $true$ because
of the existence of the path ($A_{14}$, \texttt{debits}, $E_{15}$, \texttt{credits}, $A_{17}$, \texttt{debits}, $E_{18}$, \texttt{credits}, $A_{19}$).
Another example is $Q2(P_{10},P_{13},$ (\texttt{knows,knows,worksFor}$)^+)$ that evaluates to false because there is no path from $P_{10}$ to $P_{13}$ satisfying the constraint.
\end{example}

\myQuery\ queries are also frequently occurring in real-world query logs, \textit{e.g.}, Wikidata Query Logs \cite{10.1145/3308558.3313472}, which is the largest repository of open-source graph queries (of the order of 500M queries). 
In particular, \myQuery\ queries often timed out in these logs \cite{10.1145/3308558.3313472} thus showing the limitations of graph query engines to efficiently evaluate them. 
Moreover, Neo4j (v4.3) \cite{neo4j} and TigerGraph (v3.3) \cite{Deutsch2019TigerGraphAN}, two of the mainstream graph data processing engines, do not yet support \myQuery\ queries in their current version. On the other hand, these systems have already identified the need to support these queries in the near future by following the developments of the Standard Graph Query Language (GQL) \cite{gql}.
\myQuery\ queries can be expressed in Gremlin supported by TinkerPop-Enabled Graph Systems \cite{TinkerPop-Sys}, \textit{e.g.}, Amazon Neptune \cite{Bebee2018AmazonNG}, in PGQL \cite{10.1145/2960414.2960421} supported by Oracle PGX \cite{7832832,273939}, and in SPARQL 1.1 (ASK query) supported by Virtuoso \cite{virtuoso} and Apache Jena \cite{jena}, among the others.
However, many of these systems cannot efficiently evaluate \myQuery\ queries yet as shown in our experimental study.

To the best of our knowledge, little research has been carried out on an index-based solution to evaluate \myQuery\ queries, whereas indexing is a desirable asset of future graph processing systems allowing to improve and predict the performances of graph queries 
\cite{SakrBVIAAAABBDV21}.
Previous plain reachability indexes have been well studied \cite{10.1145/67544.66950,10.1145/99935.99944,10.5555/545381.545503,1617443,4497498,10.1145/1376616.1376677,10.1145/1559845.1559930,10.1145/2463676.2465286,10.14778/2556549.2556578,10.5555/1083592.1083651,10.1145/1247480.1247573,10.14778/1920841.1920879,6544893,Veloso2014ReachabilityQI,10.14778/2732977.2732992,7750623}, but are not usable to evaluate \myQuery\ queries because of the lack of edge label information. 
More recent indexing methods have been focused on alternation of edge labels \cite{10.1145/1807167.1807183,ZOU201447,10.1145/3035918.3035955,10.14778/3380750.3380753,10.1145/3451159}, instead of concatenation in \myQuery\ queries. 

In this work, we propose the \myIndex\ index, the first reachability index tailored for \myQuery\ queries.
In the \myIndex\ index, we assign pairs of vertices and succinct label sequences to each vertex. Then, the \myIndex\ index processes an \myQuery\ query by checking whether the source vertex can reach an intermediate vertex that can also reach the target vertex under the path constraint. 

\textit{Challenge C1.} One of the challenges for building an index to process \myQuery\ queries is that there can be infinite sequences of edge labels from vertex $s$ to vertex $t$ due to the presence of cycles on paths from $s$ to $t$.
Thus, building an index for arbitrary \myQuery\ queries can be hard since the number of label sequences for cyclic graphs is exponentially growing and potentially infinite, which leads to unrealistic indexing time and index size.
We overcame this issue by leveraging a practical observation: 
the number of edge labels concatenated under the Kleene plus (or star) is typically bounded in real-world query logs, \textit{e.g.}, \cite{10.1145/3308558.3313472}. 
We refer to the maximum number of concatenated labels in a workload of  \myQuery\ queries as recursive $k$.
Given an arbitrary  recursive $k$, we show that the \myIndex\ index can be correctly built to evaluate any \myQuery\ query with a recursive concatenation of \textit{at most} $k$ arbitrary edge labels.
Note that recursive $k$ only depends on the number of concatenated labels and not on the actual length of any path selected by \myQuery\ queries.

\textit{Challenge C2.}
Although abundant indexing algorithms \cite{10.1145/67544.66950,10.1145/99935.99944,10.5555/545381.545503,1617443,4497498,10.1145/1376616.1376677,10.1145/1559845.1559930,10.1145/2463676.2465286,10.14778/2556549.2556578,10.5555/1083592.1083651,10.1145/1247480.1247573,10.14778/1920841.1920879,6544893,Veloso2014ReachabilityQI,10.14778/2732977.2732992,7750623, 10.1145/2588555.2612181, 10.1145/1871437.1871457,10.1145/2505515.2505724} have been proposed for plain reachability, which are then extended to alternation-based reachability queries \cite{10.1145/1807167.1807183,ZOU201447,10.1145/3035918.3035955,10.14778/3380750.3380753,10.1145/3451159}, none of them is able to build an index for \myQuery\ queries because of the completely different path constraints.
More precisely, indexing reachability with recursive concatenation-based path-constraints requires recording \textit{sequences} of edge labels. However, indexing reachability with alternation-based path-constraints requires \textit{sets} of edge labels, or plain reachability without edge labels. Such a fundamental difference illustrates the hardness of the problem and shows the inadequacy of existing indexing algorithms to build the \myIndex\ index.
We design a novel indexing algorithm that builds the \myIndex\ index efficiently over large and highly cyclic graphs with millions of vertices and edges as shown in our experiments.
The indexing algorithm searches and records paths with recursive concatenation-based constraints, and during each searching step, identifies the condition that allows to prune search space as well as avoid recording redundant index entries.

\textit{Contribution.}  
Our contributions are summarized as follows:
\begin{itemize}
    \item 
    We study the problem of indexing for \myQuery\ queries, \textit{i.e.}, reachability queries with a complex path constraint consisting of the Kleene plus over a concatenation of edge labels. We propose a novel instantiation of the problem to address challenge C1.

    \item We propose the \myIndex\ index, the first reachability index for processing \myQuery\ queries, along with its indexing algorithm to address challenge C2. 
    We prove that the latter builds a sound and complete \myIndex\ index for an arbitrary recursive $k$, where redundant index entries can be greedily removed.
    
    \item Our comprehensive experiments using highly cyclic real-world graphs show that the \myIndex\ index can be efficiently built and can significantly reduce the memory overhead of the (extended) transitive closure. Moreover, the \myIndex\ index is capable of answering \myQuery\ queries efficiently, up to six orders of magnitude over online traversals. We also demonstrate the speed-up and the generality of using the \myIndex\ index to accelerate query processing on mainstream graph engines. Our code is available as open source\footnote{Open Source Link: \url{https://github.com/g-rpqs/rlc-index}}.
\end{itemize}

The rest of the paper is organized as follows. Section \ref{section: related works} presents the related work while Section \ref{section: problem statement} introduces the \myQuery\ queries.
Section \ref{section: theoretical foundations} presents the theoretical foundation  and Section \ref{section: my index} describes the \myIndex\ index and its indexing algorithm.
Section \ref{section: experimental results} presents the experimental assessment of \myQuery\ queries using the  \myIndex\ index.
Section \ref{section: conclusion} concludes our work.
All related proofs are included in our online technical report \cite{2203.08606} due to the space constraint.

%% file: sections/Related_Works.tex
\section{RELATED WORK}\label{section: related works}
The two most fundamental graph queries \cite{10.1145/3104031} are graph pattern matching queries and   regular path queries (aka navigational queries). The former matches a query graph against a graph database while the latter recursively navigates a graph according to a regular expression. The two fundamental queries are completely orthogonal, and \myQuery\ queries belong to the class of the latter. 
Thus, we discuss regular path queries and related indexes in this section. 
We note approaches \cite{Ullmann1976AnAF,1323804,Madarasi2018VF2,7907163,10.14778/1453856.1453899,10.1145/1516360.1516384,10.14778/1920841.1920887,10.1145/1376616.1376660,10.1145/2463676.2465300,10.14778/2735479.2735493,10.1145/2882903.2915236,10.1007/s10115-016-0968-2,10.1145/3299869.3300086,10.1145/3299869.3319880,9035407,10.14778/3425879.3425888,10.14778/3397230.3397241} for efficient evaluating graph pattern matching queries. We refer readers to surveys \cite{10.14778/2535568.2448946,10.1145/3318464.3380581}. We do not elaborate on them further and do not consider them in our experimental evaluation due to the completely different query types.

\noindent {\bf Plain Reachability Index.} Given an unlabeled graph $G=(V,E)$ and a pair of vertices $(s,t)$, a plain reachability query asks whether there exists a path from $s$ to $t$. 
The existing approaches lie between two extremes, \textit{i.e.}, online traversals and the transitive closure.
Various indexes have been proposed. Comprehensive surveys can be found in \cite{yu2010graph,7750623,10.5555/3307192}
Plain reachability indexes mainly fall into two categories \cite{10.14778/2732977.2732992,7750623}: \textit{(1)} index-only approaches, \textit{e.g.}, \textit{Chain Cover} \cite{10.1145/99935.99944,4497498}, \textit{Tree Cover} \cite{10.1145/67544.66950}, \textit{Dual Labeling} \cite {1617443},  \textit{Path-Tree Cover} \cite{10.1145/1376616.1376677}, 2-Hop labeling \cite{10.5555/545381.545503}, TFL \cite{10.1145/2463676.2465286}, TOL \cite{10.1145/2588555.2612181}, PLL \cite{10.1145/2505515.2505724}, 3-Hop labeling \cite{10.1145/1559845.1559930}, among others; 
\textit{(2)} index-with-graph-traversal approaches, such as \textit{Tree+SSPI} \cite{10.5555/1083592.1083651}, \textit{GRIPP} \cite{10.1145/1247480.1247573}, \textit{GRAIL} \cite{10.14778/1920841.1920879},  \textit{Ferrari} \cite{6544893}, \textit{IP} \cite{10.14778/2732977.2732992}, and \textit{BFL} \cite{7750623}.

\myQuery\ queries are different from plain reachability queries because they are evaluated on labeled graphs to find the existence of a path satisfying  additional recursive label-concatenated constraints. 
Thus, the indexes used to evaluate plain reachability queries, \textit{e.g.}, 2-hop labeling \cite{10.5555/545381.545503}, are not suitable for \myQuery\ queries.
More precisely, indexing techniques for plain reachability queries only record information about graph structure but ignore information of edge labels. 

\noindent {\bf Alternation-Based Reachability Index}.
Reachability queries with a path constraint that is based on alternation of edge labels (instead of concatenation as in our work), are known as LCR queries in the literature. 
Index-based solutions for LCR queries have been extensively studied in the last decade. 

Jin \textit{et. al} \cite{10.1145/1807167.1807183} presented the first result on LCR queries. 
To compress the generalized transitive closure that records reachable pairs and sets of path-labels, the authors proposed sufficient label sets and a spanning tree with partial transitive closure. 
The Zou \textit{et. al} \cite{ZOU201447} method finds all strongly connected components (SCCs) in an input graph and replaces each SCC with a bipartite graph to obtain an edge labeled DAG. 
Valstar \textit{et. al} \cite{10.1145/3035918.3035955} proposed a landmark-based index, where the generalized transitive closure for a set of high-degree vertices called landmarks are built and an online traversal is applied to answer LCR queries, which is accelerated by hitting landmarks. 
The state-of-the-art indexing techniques for LCR queries are the Peng \textit{et. al} \cite{10.14778/3380750.3380753} method and the Chen \textit{et. al} \cite{10.1145/3451159} method.
Peng \textit{et. al} \cite{10.14778/3380750.3380753} proposed the LC 2-hop labeling, which extends the 2-hop labeling framework through adding minimal sets of path-labels for each entry in the 2-hop labeling. 
Chen \textit{et. al} \cite{10.1145/3451159} proposed a recursive method to handle LCR queries, where an input graph is recursively decomposed into spanning trees and graph summaries.

LCR queries and \myQuery\ queries have completely different regular expressions as path constraints, which makes the corresponding indexing problem inherently different.
The expression in LCR queries is an alternation of edge labels, while the one in \myQuery\ queries is a concatenation of edge labels.
The completely different path constraint makes LCR indexes inapplicable for processing \myQuery\ queries. 
The difference on path constraints also makes the indexing algorithms for LCR queries and \myQuery\ queries fundamentally different.
Specifically, for an LCR indexing algorithm, it is sufficient to traverse any cycle in a graph only once.
Conversely, in the case of \myQuery\ queries, a cycle, especially a self loop, might need to be traversed multiple times depending on label sequences along paths.
Therefore, indexing approaches for LCR queries are not applicable to indexing \myQuery\ queries.

\noindent {\bf Regular Path Queries.}
Regular path queries correspond to queries generating node pairs according to path constraints specified using regular expressions.
Under the \textit{simple} path semantics (non-repeated vertices or edges in a path), it is NP-complete to check the existence of a path satisfying a regular expression \cite{10.5555/88830.88850}. Thus, a recent work \cite{10.1145/3299869.3319882} focused on an approximate solution for evaluating regular simple path queries.
By restricting regular expressions or graph instances, there exist tractable cases \cite{10.5555/88830.88850,10.1145/2463664.2467795}, including LCR queries but not \myQuery\ queries.
When it comes to the \textit{arbitrary} path semantics (allowing repeated vertices or edges in a path), regular path queries for generating node pairs can be processed by using automata-based techniques \cite{10.1145/2206869.2206879,10.1145/2463664.2465216} or bi-directional BFSs from rare labels \cite{10.1007/978-3-642-31235-9_12}.
Optimal solutions can be used for sub-classes of regular expressions, \textit{e.g.}, a matrix-based method \cite{5767858} for the case without recursive concatenation, and a B+tree index for the case with a bounded path length (non-recursive path constraints) \cite{Fletcher2016EfficientRP} and a fixed path pattern \cite{Kuijpers2021PathII}.
There also exist in the literature partial evaluation for distributed graphs \cite{10.14778/2350229.2350248}, and incremental approaches for streaming graphs \cite{10.1145/3318464.3389733,pacaci2021evaluating}.
Compared to all these works, we focus on designing an index-based solution to handle reachability queries with path constraints of recursive concatenation  over a static and centralized graph under the arbitrary path semantics, which is an open challenge in the design of reachability indexes.
To the best of our knowledge, our work is the first of its kind focusing on the design of a reachability index for such queries.

%% file: sections/Preliminary.tex
\section{PROBLEM STATEMENT}\label{section: problem statement}
An edge-labeled graph is $G=(V,E,\Lab)$, where $\Lab$ is a finite set of labels, $V$ a finite set of vertices,  and $E\subseteq V \times \Lab \times V$ a finite set of labeled edges.
For the graph in Fig. \ref{fig: example in introduction}, we have $\Lab=\{$\texttt{knows, worksFor, debits, credits, holds}$\}$, and that $e_1$=($P_{10}$, \texttt{knows}, $P_{11}$) is a labeled edge.
We use $\lambda: E \rightarrow \Lab$ to denote the mapping from an edge to its label, \textit{e.g.},  $\lambda(e_1)= \texttt{knows}$.
The frequently used symbols are summarized in Table \ref{table: notations}.

In this paper, we consider the \textit{arbitrary path} semantics \cite{10.1145/3104031} for evaluating \myQuery\ queries, \textit{i.e.}, vertices or edges can appear more than once along the path. The arbitrary path semantics is widely adopted in practical graph query languages, \textit{e.g.}, SPARQL 1.1 and PGQL, because evaluating a reachability query with an arbitrary regular expression as the path constraint is tractable under the arbitrary path semantics but is NP-complete under the simple path semantics, \textit{e.g.}, \myQuery\ queries. It should be noteworthy that LCR queries with alternation-based path constraints belong to a special class (\textit{trC} class \cite{10.1145/2463664.2467795}), which remain tractable under the simple path semantics.

An arbitrary path $p$ in $G$ is a vertex-edge alternating sequence $p(v_0,v_n)=(v_0,e_1,...,e_n,v_n)$, where $n\geq 1$, and $v_0,...,v_n\in V$, $e_1,...,e_n\in E$, and $|p(v_0,v_n)|=n$ that is the length of the path. 
For $p(v_0,v_n)$, $v_0$ is the source vertex and $v_n$ is the target vertex.
If there exists a path from $v_0$ to $v_n$, then $v_0$ reaches $v_n$, denoted as $v_0\leadsto v_n$.
The label sequence of the path $p(v_0,v_n)$ is $\Lambda(p(v_0,v_n))=(\lambda(e_1),...,\lambda(e_n))$. 
When the context is clear, we also use $\Lambda(u,v)$ to denote the sequence of edge labels of a path from $u$ to $v$.

\subsection{Minimum Repeats}
We use $l_i$ to denote an edge label, $L=(l_1,...,l_{n})$ a label sequence, $|L|=n$ the length of $L$, and $\epsilon$ the empty label sequence, \textit{i.e.}, $|\epsilon|=0$.
We use $\circ$ to denote the concatenation of label sequences (or labels), \textit{i.e.}, $(l_1,...,l_i)\circ(l_{i+1},...,l_n)=(l_{1},...,l_n)$,
or $L\circ L = L^2$.
For the empty label sequence $\epsilon$, we define $L\circ \epsilon = \epsilon \circ L = L$.

The sequence $L'=(l'_1,...,l'_{n'}), |L'|=n'$ is a \textit{repeat} of $L=(l_1,...,l_{n})$, if there exists an integer $z$, such that $\frac{n}{n'}=z\geq 1$, and $l'_j=l_{j+i\times n'}$ for every $j\in(1,...,n')$ and $i\in(0,...,z-1)$.
A repeat $L'$ of $L$ is minimum if $L'$ has the shortest length of all the repeats of $L$.
The \textit{minimum repeat} (MR) of  $L$ is denoted as $MR(L)$ that is also a sequence of edge labels.
For example, given the path $p=(P_{10}$, \texttt{knows},  $P_{11}$, \texttt{worksFor}, $P_{12}$, \texttt{knows}, $P_{13}$, \texttt{worksFor}, $P_{16})$ in Fig. \ref{fig: example in introduction}, we have $MR(p(P_{10},P_{16}))= (\texttt{knows,worksFor})$.
If $L=MR(L)$, we also say $L$ itself is a minimum repeat.
Given a positive integer $k$ and a label sequence $L$, if $|MR(L)|\leq k$, then we say $L$ has a non-empty k-MR that is $MR(L)$.

\begin{lemma}\label{lemma: unique MR}
For a label sequence $L$, $MR(L)$ is unique.
\end{lemma}
Suppose $L$ has two minimum repeats, then only the shorter is $MR(L)$. Therefore, we have Lemma \ref{lemma: unique MR}.

\subsection{\myQuery\ Query}\label{definition: my query}
We consider the path constraint $L^+=(l_1,..., l_{k})^+$, where `+' is the Kleene plus, \textit{i.e.}, one-or-more concatenations of the sequence $L=(l_1,..., l_{k})$.
A label sequence $\Lambda(u,v)$ of a path $p(u,v)$ satisfies a label-constraint $L^+$, if and only if $MR(\Lambda(u,v))=L$.
If such a path $p(u,v)$ exists, then $u$ can reach $v$ with the constraint $L^+$, denoted as
$u \overset{L^+}{\leadsto}  v$, otherwise $u \overset{L^+}{\not\leadsto}  v$.
Notice that reachability queries with path-constraints based on recursive concatenation could ask for additional constraints related to path length. For instance, if the path constraint is $L^+=(\texttt{knows, knows})^+$, then the query would need to additionally check whether the path length is even.
In general, such fragment of queries, \textit{i.e.}, queries with $L^+$ s.t. $L \neq MR(L)$, impose an additional constraint on path length leading to the complicated even-path problem (NP-complete \cite{LaPaugh1984TheEP,10.5555/88830.88850,10.1145/2463664.2467795} for simple paths), which is beyond the scope of our paper.

\begin{definition}[RLC Query]
Given an edge-labeled directed graph $G=(V,E,\Lab)$ and a recursive $k$, an \myQuery\ query is a triple $(s,t,L^+)$, $L=(l_1,...,l_j)$, where $s,t\in V$, $L=MR(L)$, $j\leq k$, and $l_i\in \Lab$ for $i\in(1,...,j)$.
If $s \overset{L^+}{\leadsto}t$, then the answer to the query is \textit{true}. Otherwise, the answer is \textit{false}.
\end{definition}

For the sake of simplicity in this paper we focus on the  \myQuery\ queries with the Kleene plus. Queries $(s,t,L^*)$ with the Kleene star can be trivially reduced to queries with the Kleene plus $(s,t,L^+)$ through checking whether $s$ is equal to $t$. Our method is thus applicable to \myQuery\ queries with the Kleene star.

Given an \myQuery\ query $Q(s,t,L^+)$, under the arbitrary path semantics, two naive approaches can be used to evaluate $Q$.
As \myQuery\ queries are path queries with regular expressions, in the first approach \myQuery\ queries can be evaluated by online graph traversals, \textit{e.g.}, BFS, guided by a minimized NFA (nondeterministic finite automaton) \cite{10.5555/3307192} that is constructed according to the regular expression in an \myQuery\ query.
The second approach leads to pre-computing the transitive closure, that for each pair of vertices $(s,t)$ records whether $s\leadsto t$ and all the label sequences from $s$ to $t$. 
Notice that the naive approach to build the transitive closure is not usable in our case because cycles may exist on the path from $s$ to $t$ such that the BFS from $s$ can generate infinite label sequences for paths to $t$.
To address this issue, we adopt an \textit{extended transitive closure}, which is presented in Section \ref{section: experimental results}.
As demonstrated in our experiments, these two solutions require either too much query time or storage space and are impractical for large graphs.

\subsection{Indexing Problem}
Our goal is to build an index to efficiently process \myQuery\ queries. The  indexing problem is summarized as follows.
\begin{problem}
Given an edge-labeled graph $G$, the indexing problem is to build a reachability index for processing \myQuery\ queries on $G$,  such that the size of the index is minimal and the correctness of query processing is preserved.
\end{problem}

We observe that recording MRs, instead of raw label sequences of paths in $G$, can reduce the storage space, and such a strategy does not violate the correctness of query processing.
The main benefits are twofold: 
(1) MRs are not longer than raw label sequences; (2) different raw label sequences may have the same MR.
For example, in Fig. \ref{fig: example in introduction}, there exist two paths from $P_{10}$ to $P_{16}$ having the label sequence
(\texttt{knows, knows, knows, knows}) and (\texttt{knows, knows, knows}), which have the same MR \texttt{knows}.

\begin{definition}[Concise Label Sequences]\label{definition: cls}
Let $\mathbb{P}(s,t)$ be the set of all paths from $s$ to $t$.
The concise set of label sequences from vertex $s$ to $t$, denoted as $S^k(s,t)$, is the set of k-MRs of all label sequences from $s$ to $t$, \textit{i.e.},
$S^k(s,t)=\{L|p\in \mathbb{P}(s,t),  L = MR(\Lambda(p)), |L|\leq k\}$.
\end{definition}

To process \myQuery\ queries, we need to compute and record the concise label sequences. 
We have Proposition \ref{proposition: cls} by definition.

\begin{proposition}\label{proposition: cls}
$s\overset{L^+}{\leadsto}t,|L|\leq k$ in $G$ iff $ L\in S^k(s,t)$.
\end{proposition}
For example, in Fig. \ref{fig: example in introduction}, we have $S^2(P_{12},P_{16})=\{$(\texttt{knows}), (\texttt{knows, worksFor})$\}$.
With $S^2(P_{12},P_{16})$, \myQuery\ queries from $P_{12}$ to $P_{16}$ can be processed correctly.

%% file: sections/Concatenation.tex
\begin{table}
    \centering
    \caption{Frequently used symbols.}\label{table: notations}
    \resizebox{\linewidth}{!}{
        \begin{tabular}{|l|l|} \hline
           \textbf{Notation}                        & \textbf{Description}                                      \\ \hline
            $p$, or $p(u,v)$                        & a path, or the path from $u$ to $v$                       \\ \hline
            $\circ$                                 & concatenation of labels or label sequences                \\ \hline
		    $\Lambda(u,v)$, or $\Lambda(p(u,v))$	& the label sequence of a path from $u$ to $v$              \\ \hline
		    $L$		                                & a label sequence                                          \\ \hline
		    $L^+$		                            & a label constraint                                        \\ \hline
		    $MR(L)$                                 & the minimum repeat of a label sequence $L$                \\  \hline
		    $k$                                     & the upper bound of the number of labels in $L^+$          \\ \hline
		    $S^k(u,v)$                              & the concise set of minimum repeats from $u$ to $v$        \\ \hline
            $u \overset{L^+}{\leadsto} v$, or 
                $u \overset{L^+}{\not\leadsto} v$
								                    & $u$ reaches $v$ through an $L^+$-path, or otherwise       \\ \hline                                
            $u \leadsto v$, or $u \not\leadsto v$     
                                                    & $u$ reaches $v$, or otherwise                             \\ \hline                               
            $in(v)$, or $out(v)$                    & the set of vertices that can reach $v$, or $v$ can reach  \\ \hline
            $aid(v)$                                & the access id of vertex $v$ by the indexing algorithm     \\ \hline
            $v^{(j)}_i$                                 & a vertex with vertex id $i$ and access id $j$             \\ \hline
        \end{tabular}
    }
    \vspace*{-0.5cm}
\end{table}

\section{Kernel-based search}\label{section: theoretical foundations}
In this section, we deal with the following question: \textit{how to compute concise label sequences?}
The problem for computing a concise label sequence is that if a cycle exists on a path from $s$ to $t$, there exist infinite paths from $s$ to $t$, which makes the computation of $S^k(s,t)$ infeasible, \textit{e.g.}, $|\mathbb{P}(P_{11},P_{13})|$ in Fig. \ref{fig: example in introduction} is infinite.
We overcome this issue by leveraging the upper bound of recursively concatenated labels in a constraint, \textit{i.e.}, recursive $k$.
We observed that we don't have to compute all possible label sequences for paths going from $P_{11}$ to $P_{13}$ as the set of label sequences $L$ such that $|MR(L)| \leq k$ is actually finite.
In general, let $v$ be an intermediate vertex that a forward breadth-first search from $s$ is visiting.
The main idea is that when the path from $s$ to $v$ reaches a specific length, we can decide whether we need to further explore the outgoing neighbours of $v$. 
Moreover, if the outgoing neighbours of $v$ are worth exploring, the following search can be guided by a specific label constraint.
In the following, we first provide an illustrating example, and then formally define the specific constraint that is used to guide the subsequent search.

\begin{example}[Illustrating Example]\label{example: KBS}
Consider the graph in Fig. \ref{fig: example in introduction}. Assume we need to compute $S^2(P_{11},P_{13})$, \textit{i.e.}, $k=2$, and we perform a breadth-first search from $P_{11}$.
When $P_{13}$ is visited for the first time, we add (\texttt{knows}) and (\texttt{worksFor, knows}) into $S^2(P_{11},P_{13})$. After that, when the search depth reaches $2k=4$, \textit{i.e.}, $P_{12}$ is visited for the second time, we can have $4$ different label sequences, which are $L_1=$ (\texttt{knows, knows, knows, knows}), $L_2=$ (\texttt{knows, knows, knows, worksFor}), $L_3=$ (\texttt{worksFor, knows, knows, knows}), and $L_4=$ (\texttt{worksFor, knows, knows, worksFor}). Given this, all the $4$ label sequences except $L_1$ do not need to be expanded anymore, because their expansions cannot produce a minimum repeat whose length is not larger than $2$.
After this, the following search continued with $L_1$ is guided by $(\texttt{knows})^+$ that is computed from $L_1$. 
However, because there already exists (\texttt{knows}) in $S^2(P_{11},P_{13})$, the search terminates.
\end{example}

\begin{definition}
If a label sequence $L$ can be represented as $L=(L')^h\circ L''$, where $h\geq 2$,
$L'\neq \epsilon$ and $MR(L')=L'$, and $L''$ is $\epsilon$ or a proper prefix of $L'$, then $L$ has the \textit{kernel} $L'$ and the \textit{tail} $L''$.       
\end{definition}

For example, the label sequence (\texttt{knows, knows, knows, knows}) from $P_{11}$ has a kernel \texttt{knows} and a tail $\epsilon$.

\textit{Kernel-based search.}
When a kernel has been determined at a vertex that is being visited, the subsequent search to compute $S^k(s,t)$ can be guided by the Kleene plus of the kernel, \textit{e.g.}, $(\texttt{knows})^+$ is used to guide the search in Example \ref{example: KBS}.
We call this approach KBS (\textit{kernel-based search}) in the remainder of this paper.
In a nutshell, KBS consists of two phases: (1) \textit{kernel-search} and (2) \textit{kernel-BFS}, where the first phase is to compute kernels, and the second to perform kernel-guided BFS. 
We show in Theorem \ref{theorem: 2k search} that KBS can compute a sound and complete $S^k(s,t)$. The proof of Theorem \ref{theorem: 2k search} is included in our technical report \cite{2203.08606} due to the space limit. 

\begin{theorem}\label{theorem: 2k search}
    Given a  path $p$ from $u$ to $v$ and a positive integer $k$, $p$ has a non-empty k-MR if and only if one of the following conditions is satisfied,
    \begin{itemize}[leftmargin = *]
        \item[-]
        Case 1: $|p|\leq k$. $MR(\Lambda(p))$ is the k-MR of $p$;
        \item[-]
        Case 2: $k<|p|\leq 2k$.  If $|MR(\Lambda(p))|\leq k$, $MR(\Lambda(p))$ is the k-MR of $p$;
        \item[-] 
        Case 3: $|p|> 2k$. Let $x$ be the intermediate vertex on $p$, s.t. $|p(u,x)|=2k$.
        If $\Lambda(p(u,x))$ has a kernel $L'$ and a tail $L''$, and $MR(L''\circ \Lambda(p(x,v)))=L'$, then $L'$ is the k-MR of $p$. 
    \end{itemize}
\end{theorem}

We discuss below two strategies to compute kernels based on Theorem \ref{theorem: 2k search}, namely \textit{lazy} KBS and \textit{eager} KBS, and explain why eager KBS is better than lazy KBS, which is used in our indexing algorithm presented in Section \ref{section: indexing algorithm}. 

\textit{Lazy KBS}. Theorem \ref{theorem: 2k search} can be transformed into an algorithm to find kernels, \textit{i.e.}, for a source vertex we generate all paths of length $2k$, and then compute all the kernels of these paths. This strategy is referred to as lazy KBS, which means kernels are correctly determined when the length of paths reaches $2k$, \textit{e.g.},  lazy KBS is used in Example \ref{example: KBS}. 

\textit{Eager KBS}. In contrast to the lazy strategy, we can determine kernel candidates earlier, instead of valid kernels that require the length of paths to be $2k$.
The main idea is to treat any k-MR that is computed using any path $p,|p|\leq k$ as a kernel candidate which is then used to guide KBS. 
Although an invalid kernel may be included, the search guided by the invalid kernel will not reach a target vertex through a path of which the k-MR is the invalid kernel. Thus, the result computed by the eager strategy is still sound and complete.  

\begin{example}
    Consider the example of computing $S^2(P_{10},P_{13})$ in Fig. \ref{fig: example in introduction}. Using the eager strategy, 
    when $P_{12}$ is visited for the first time, two kernel candidates can be determined, \textit{i.e.}, \texttt{(knows)} and \texttt{(knows, worksFor)}. Although $(\texttt{knows, worksFor})^+$ is an invalid kernel, the search guided by it cannot reach $P_{13}$.
\end{example}
    
The key advantage of the eager strategy over the lazy strategy is that it allows us to advance KBS from the kernel-search phase to the kernel-BFS phase.
This can make KBS more efficient because generating all label sequences of length $2k$ from a source vertex is more expensive than the case of  paths of length $k$, especially on a dense graph. 

%% file: sections/MyIndex.tex
\section{\myIndex\ Index}\label{section: my index}
In this section, we present the \myIndex\ index, and the corresponding query and indexing algorithm.

\subsection{Overarching Idea}\label{section: index definition}
Given an \myQuery\ query $(s,t,L^+), |L|\leq k$, the idea is to check whether there exists a path $(s,...,u,...,t)$ whose label sequence satisfies the label constraint $L^+$, where $u$ is an intermediate vertex in $p$.
In other words, the query is answered by concatenating two MRs of the sub-paths of $p$, \textit{i.e.}, $MR(\Lambda(s,u))$ and $MR(\Lambda(u,t))$.

\begin{definition}[RLC Index]\label{definition: myIndex}
Let $G=(V,E,\Lab)$ be an edge-labeled graph and recursive $k$ be a positive integer. The \myIndex\ index of $G$ assigns to each vertex $v\in V$ two sets:
$\lin(v)=\{(u,L')|  u  \leadsto v, L'\in S^k(u,v))\}$, and $\lout(v)=\{(w,L'')|  v \leadsto  w , L''\in S^k(v,w)\}$.
Therefore, there is a path $p(s,t)$ satisfying an arbitrary constraint $L^+, |L|\leq k$, if and only if one of the following cases is satisfied,
\begin{itemize}[leftmargin = *]
    \item Case 1: $\exists (x,L')\in \lout(s)$ and $\exists (x,L'')\in \lin(t)$,  such that $L'= L''=L$;
    \item Case 2: $\exists (t,L''')\in \lout(s)$ or $\exists (s,L''')\in \lin(t)$, such that $L'''=L$. 
\end{itemize}
\end{definition}

\begin{figure}
	\centering
\begin{tikzpicture}[node distance=18mm, thick, node/.style = {draw, scale=1, circle,inner sep=2,outer sep=2}, label/.style = {scale = 1, sloped, midway, above}] 
	\node[node] (1) {$v^{{(1)}}_{1}$};
	\node[node] (2) [below right of=1]{$v^{{(3)}}_{2}$};
	\node[node] (3) [below left of=1]{$v^{{(2)}}_{3}$};
	\node[node] (4) [above left of=3]{$v^{{(4)}}_{4}$};
	\node[node] (5) [above right of=2]{$v^{{(5)}}_{5}$};
	\node[node] (6) [below left of=4]{$v^{{(6)}}_{6}$};
	\draw[->] (5) -- node[label] {$l_1$}(1); 
	\draw[->] (2) to [out=60,in=270-60,looseness=1] node[label] {$l_1$}(5); 
	\draw[->] (1) -- node[label] {$l_1$}(2); 
	\draw[->] (1) to [out=180+30,in=90-30,looseness=1] node[label] {$l_2$}(3);
	\draw[->] (3) -- node[label][pos = 0.6] {$l_1$}(2);
	\draw[->] (3) -- node[label] {$l_2$}(4);
	\draw[->] (4) -- node[label] {$l_3$}(6);
	\draw[->] (4) -- node[label][pos = 0.5] {$l_1$}(1);
	\draw[->] (3) -- node[label] {$l_1$}(6);
	\draw[->] (3) to [out=30,in=270-30,looseness=1] node[label][below][pos = 0.6] {$l_2$}(1);
	\draw[->] (2) to [out=30,in=270-30,looseness=1] node[label][below][pos = 0.6] {$l_2$}(5);
\end{tikzpicture} 
	\caption{A graph instance G for illustrating the RLC index.} \label{fig: running example of my index}
	\vspace*{-0.5cm}
\end{figure}
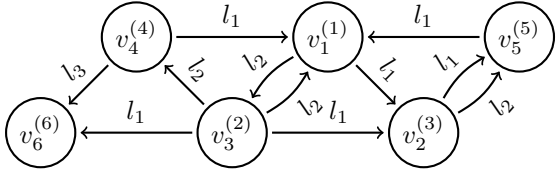

\begin{example}[\textit{Running Example of the RLC Index}]
Consider the graph $G$ shown in Fig. \ref{fig: running example of my index}. 
The \myIndex\ index with recursive $k=2$ for $G$ is presented in Table \ref{table: final my index}.
We have $Q_1(v_{3},v_{6},(l_2,l_1)^+)=true$ because $\exists (v_1,(l_2,l_1))\in \lout(v_3)$ and $\exists (v_{1},(l_2,l_1))\in \lin(v_{6})$.
Indeed, there exists the path $(v_3,l_2,v_4,l_1,v_1,l_2,v_3,l_1,v_6)$ from $v_3$ to $v_6$ in the graph in Fig. \ref{fig: running example of my index}.
For $Q_2(v_{1},v_{2},(l_2,l_1)^+)$, the answer is $true$ because $\exists (v_{1},(l_2,l_1))\in \lin(v_{2})$.
Given $Q_3(v_1,v_3,(l_1)^+)$, the answer is $false$. Although $v_1$ can reach $v_3$, \textit{e.g.}, $\exists (v_1,l_2)\in \lin(v_3)$, the constraint $(l_1)^+$ of $Q_3$ cannot be satisfied. 
\end{example}

Whereas our indexing framework leverages the canonical 2-hop labeling
framework for plain reachability queries \cite{10.5555/545381.545503}, indexing \myQuery\ queries is inherently more challenging due to the presence of complex  recursive label concatenations, which calls for the design of a novel indexing algorithm.

In order to save storage space, when building an \myIndex\ index,  redundant index entries have to be removed as many as possible, \textit{i.e.}, building a minimal (or condensed) \myIndex\ index.
The underlying idea is that if there exists a path $p$ such that $u \overset{L^+}{\leadsto} v$, then the \myIndex\ index only records the reachability information of the path $p$ once, \textit{i.e.}, either through Case 1 or Case 2 in Definition \ref{definition: myIndex}. This leads to the following definition. 

\begin{definition}[Condensed RLC Index]
    The \myIndex\ index is condensed, if for every index entry $(s,L)\in \lin(t)$ (or $(t,L)\in \lout(s)$), there do not exist index entries $(u,L')\in \lout(s)$ and $(u,L'')\in \lin(t)$ such that $L = L' =L ''$.
\end{definition}

We focus on designing an indexing algorithm that can build a correct (sound and complete) and condensed \myIndex\ index.

\subsection{Query and Indexing Algorithm} \label{section: indexing algorithm}
The query algorithm is presented in Algorithm \ref{algo: query algorithm}, where we use $I$ to denote an index entry.
Each index entry $I$ has the schema $(vid,mr)$, where $vid$ represents vertex id and  $mr$ recorded minimal repeat.
Given an \myQuery\ query $(s,t,L^+)$, to efficiently find $(u,L')\in \lout(s)$ and $(u,L'')\in \lin(t)$, we execute a merge join over $\lout(s)$ and $\lin(t)$, shown at line 4 in Algorithm \ref{algo: query algorithm}.
The output of the merge join is a set of index entry pairs $(I',I'')$, s.t. $I'.vid = I''.vid$.
Case 1 of the \myIndex\ index (see Definition \ref{definition: myIndex}) is checked at line 5. Case 2 is checked at line 3.
If one of these cases can be satisfied, the  answer \textit{true} will be returned immediately. Otherwise, index entries in $\lout(s)$ and $\lin(t)$  are exhaustively merged, and the answer \textit{false} will be returned at last.

\begin{table}
	\centering	
	\caption{The \myIndex\ index for the graph in Fig. \ref{fig: running example of my index}.}\label{table: final my index}
	\resizebox{\linewidth}{!}{
	\begin{tabular}{|c|l|l|}
				\hline
				\textbf{V} 	    &  $\bm{\lin(v)}$                                   & $\bm{\lout(v)}$ 		                        \\ \hline \hline
				$v_{1}$			& -											        & $(v_1,l_2)$, $(v_1,l_1)$,	$(v_1,(l_2,l_1))$	\\ \hline
				$v_{2}$			& $(v_1,l_1)$, $(v_1,(l_2,l_1))$                    & $(v_1,(l_2,l_1))$, $(v_1,l_1)$				\\ \hline
				$v_{3}$			& $(v_1, l_2)$, $(v_1, (l_1,l_2))$					& \begin{tabular}{@{}l@{}} $(v_1,l_2)$,
				                                                                        $(v_1,(l_2,l_1))$, $(v_1,l_1)$, \\ 
				                                                                        $(v_3, (l_1,l_2))$ \end{tabular}           \\ \hline
				$v_{4}$			& $(v_1, l_2)$               				        & $(v_1,l_1)$, $(v_3,(l_1,l_2))$ 	           \\ \hline
				$v_{5}$			& \begin{tabular}{@{}l@{}}$(v_1,(l_1,l_2))$, 
				                    $(v_1,l_1)$,$(v_3,(l_1,l_2))$,\\$(v_2, l_2)$    
				                    \end{tabular}                                   & $(v_1,l_1)$, $(v_3,(l_1,l_2))$                \\ \hline
				$v_{6}$			& \begin{tabular}{@{}l@{}}
				                    $(v_{1},(l_2,l_1))$, $(v_{3},l_1)$, 
				                    $(v_{3},(l_2,l_3))$,\\ 
				                    $(v_4,l_3)$    
				                  \end{tabular}                                     & - 	 					                    \\ \hline
	\end{tabular}
	}
\end{table}

In the remainder of this subsection, we present an indexing algorithm (Algorithm \ref{algo: indexing algorithm}) to build the \myIndex\ index that is sound, complete and condensed. 
We use $v_i$ to denote a vertex with id $i$.
Given a graph $G(V,E,\Lab)$, the indexing algorithm mainly performs backward and forward KBS from each vertex in $V$ to create index entries, and \textit{pruning rules} are applied to accelerate index building as well as remove redundant index entries.

\textit{Indexing using KBS}.
We explain below how the backward KBS creates $\lout$-entries. 
The forward KBS follows the same procedure, except that $\lin$-entries will be created. 
Suppose that the backward KBS from vertex $v_i$ is visiting $v$.
If $|MR(\Lambda(v,v_{i}))|\leq k$, then we add $(v_{i},MR(\Lambda(v,v_{i})))$ into $\lout(v)$.
Although there may be cycles in a graph, the KBS will not go on forever, because when the depth of the search reaches $k$, the KBS will be transformed  from its kernel-search phase into its kernel-BFS phase that is then guided by the Kleene plus of a kernel candidate, such that the KBS can terminate if any invalid label (or state) transition is met, or a vertex being visited with a label $l_i$ of the kernel has been already visited with a label $l_j$ of the kernel, s.t. $i=j$.

KBSs are executed from each vertex in $V$ and the execution follows a specific order. 
The idea is to start with vertices that have more connections to others, allowing such vertices to be intermediate hops to remove redundancy.
In the \myIndex\ index, we leverage the IN-OUT strategy, \textit{i.e.}, sorting vertices according to $(|out(v)| + 1) \times 
(|in(v)| + 1)$ in descending order, which is known as an efficient and effective strategy for various reachability indexes based on the 2-hop labeling framework.
The id of vertex $v$ in the sorted list is referred to as the access id of $v$, denoted as $aid(v)$ starting from $1$, \textit{e.g.}, for the graph in Fig. 2, the sorted list is $(v_1,v_3,v_2,v_4,v_5,v_6)$, where $aid(v_3)=2$, or simply $v^{(2)}_3$ in Fig. \ref{fig: running example of my index}.

\begin{algorithm}[t]
\footnotesize
\SetAlgoVlined
\SetKwFunction{KwQuery}{Query}
\SetKwProg{Fn}{procedure}{}{}
\Fn{\KwQuery{$s,t,L^+$}}{
    \If{$\exists (t,L)\in \lout(s)$ {\bf or} $\exists (s,L)\in \lin(t)$}{
        \KwRet{true}\;
    }
    \For{$(I',I'')\in \texttt{mergeJoin}( \lout(s), \lin(t))$}{
        \If{$I'.mr = L$ {\bf and} $I''.mr = L$}{
            \KwRet{true}\;
        }
    }
    \KwRet{false}\;
}
\caption{Query Algorithm.}\label{algo: query algorithm}
\end{algorithm}
\begin{algorithm}[t]
\footnotesize
\SetAlgoVlined
\SetKwFunction{KBS}{\texttt{kernelBasedSearch}}
\SetKwFunction{KwFBFS}{\texttt{forwardKernelSearch}}
\SetKwFunction{KwBBFS}{\texttt{backwardKernelSearch}}
\SetKwFunction{KwCreateEntries}{\texttt{insert}}
\SetKwFunction{KwBBFSWithKernel}{\texttt{backwardKernelBFS}}
\SetKwFunction{KwFBFSWithKernel}{\texttt{forwardKernelBFS}}
\SetKwProg{Fn}{procedure}{}{}
\newcommand{\forcond}{$i=0$ \KwTo $n-1$}
\Fn{\KBS{$v,k$}}{
     \For{ $(L, vSet)\in$ \KwBBFS{$v,k$}}{
        \KwBBFSWithKernel{$v,vSet,L$}\;
     }
    \For{ $(L, vSet)\in$ \KwFBFS{$v,k$}}{
        \KwFBFSWithKernel{$v,vSet,L$}\;
    }
}
\Fn{\KwBBFS{$v,k$}}{
    $q\gets $ an empty queue of (vertex, label sequence)\;
    $q$.enqueue($v,\epsilon$)\;
    $map \gets $ a map of (kernel candidates, vertex set)\;
    \While{$q$ is not empty}{
        $(x,seq)\gets q$.dequeue()\;
        \For{in-coming edge $e(y,x)$ to $x$}{
            $seq'\gets \lambda(e(y,x)) \circ seq$;
            $L\gets $ MR($seq'$)\;
            \KwCreateEntries($y,v,L$)\; 
            $map.get(L).add(x)$\;
            \If{$|seq'|<k$}{
                $q$.enqueue($y,seq'$)\;
            }
        }
    }
    \KwRet{$map$}\;
}
\Fn{\KwCreateEntries{$s,t,L$}}{
    \If{$aid(t) > aid(s)$ {\bf or} Query($s,t,L^+$)}{
        \KwRet{false}\;
    }\Else{
        add $(t,L)$ into $\lout(s)$\;
        \KwRet{true}\;
    }
}

\Fn{\KwBBFSWithKernel{$v,vSet,L$}}{
    $q \gets$ an empty queue of $(vertex, integer)$\;
    \For{$x\in vSet$}{
        mark $x$ as visited by state $1$, $q$.enqueue($x,|L|$)\;
    }
    \While{$q$ is not empty}{
     $(x,i)\gets q$.dequeue(), $i \gets i-1$\;
        \lIf{$i=0$}{
            $i=|L|$
         }
         label $l\gets L.get(i)$\;
        \For{in-coming edge $e(y,x)$ to $x$}{
            \If{$l\neq \lambda(e(y,x))$ {\bf{or}} $y$ was visited by state $i$}{
                \bf{continue}\;
            }
            \If{{ $i=1$ \bf{and}} $\KwCreateEntries(y,v,L)$}{
                \bf{continue}\;
            }
            $q$.enqueue($y,i$); mark $y$ visited by state $i$\;
        }
    }
}
\caption{Indexing Algorithm.}\label{algo: indexing algorithm}
\end{algorithm}

\begin{example}[\textit{Running Example of Indexing}]
Consider the graph in Fig. \ref{fig: running example of my index} and the \myIndex\ index in Table \ref{table: final my index} with recursive $k=2$.
The KBSs are executed from each vertex in the order of $(v_1,v_3,v_2,v_4,v_5,v_6)$.
We explain below the backward KBS from $v_1$, which is the first search of the indexing algorithm. 
The traversal of depth $1$ of this backward KBS visits $v_4$ and creates $(v_1,l_1)$ in $\lout(v_4)$, visits $v_3$ and creates $(v_1,l_2)$ in $\lout(v_3)$, and visits $v_5$ and creates $(v_1,l_1)\in \lout(v_5)$.
The traversal of depth $2$ creates $(v_1,(l_2,l_1))$ in $\lout(v_3)$, $(v_1,l_2)$ in $\lout(v_1)$, $(v_1,l_1)\in \lout(v_2)$, and  $(v_1,(l_2,l_1))\in \lout(v_2)$.
Then, the kernel-search phase of this KBS terminates as the depth of the search reaches $2$, which generates kernel candidate $l_1$ with a set of frontier vertices $\{v_4, v_5, v_2\}$, kernel candidate $l_2$ with a set of  frontier vertices $\{v_3,v_1\}$, and kernel candidate $(l_2,l_1)$ with a set of frontier vertices $(v_3, v_2)$.
After this, this KBS is turned into three kernel-BFSs guided by $(l_1)^+$, $(l_2)^+$, and $(l_2,l_1)^+$ with the corresponding frontier vertices.
The kernel-BFS terminates under the case of an invalid label transition or a repeated visiting.
For example, the label of the incoming edge of $v_3$ is $l_2$, which is an invalid state transition of $(l_2,l_1)^+$ in the backward kernel-BFS from $v_1$, such that the kernel-BFS guided by $(l_2,l_1)^+$ terminates at $v_3$.
For another example, index entry $(v_1,l_1)\in \lout(v_1)$ is created when $v_1$ is visited for the first time by the kernel-BFS from $v_1$ guided by $(l_1)^+$, but this kernel-BFS will not continue when it visits $v_5$ that has already been visited with the label $l_1$.
\end{example}

\textit{Pruning Rules}.
To speed up index construction and remove redundant index entries, we apply pruning rules during KBSs. 
For ease of presentation, we present the rules for backward KBSs, and the same rules apply for forward ones.
\begin{itemize}[leftmargin = *]
    \item[-] \textit{\textbf{PR1:} If the k-MR of an index entry that needs to be recorded can be acquired from the current snapshot of the \myIndex\ index, then the index entry can be skipped.}
    
    \item[-] \textit{\textbf{PR2:} If vertex $v_{i}$ is visited by the backward KBS performed from vertex $v_{i'}$ s.t. $aid(v_{i'})>aid(v_i)$, then the corresponding index entry can be skipped.}

    \item[-] \textit{\textbf{PR3:} If vertex $v_{i}$ is visited by the  kernel-BFS phase of a backward KBS performed from vertex $v_{i'}$, and PR1 (or PR2) is triggered, then vertex $v_{i}$ and all the vertices in $in(v_i)$ are skipped.}
\end{itemize}
The correctness of Algorithm\ \ref{algo: indexing algorithm} with pruning rules is guaranteed by Theorem \ref{theorem: correctness of myIndex} presented in Section \ref{subsection: analysis}.

\begin{example}[\textit{Running Example of Pruning Rules}]
Consider the forward KBS from $v_3$ for the graph in Fig. \ref{fig: running example of my index}.
It can visit $v_2$ through label sequence $(l_2,l_1)$, such that it tries to create $(v_3,(l_2,l_1))$ in $\lin(v_2)$.
However, there already exist $(v_1,(l_2,l_1))\in \lout(v_3)$ and $(v_1,(l_2,l_1))\in \lin(v_2)$, such that $Q(v_3,v_2,(l_2,l_1)^+)=true$ with the current snapshot of the \myIndex\ index, \textit{i.e.}, the reachability information has already been recorded. 
Thus, the index entry $(v_3,(l_2,l_1))$ in $\lin(v_2)$ is pruned according to PR1. 
As an example of PR2, consider the backward KBS from $v_2$.
It can visit $v_1$ through path $(v_1,l_2,v_3,l_1,v_2)$, such that it tries to create $(v_2,(l_2,l_1))$ in $\lout(v_1)$. 
Given $aid(v_2) > aid(v_1)$, such that the index entry can be pruned by PR2.
Consider the forward KBS from $v_2$ for an example of PR3. 
It visits $v_2$ through path $(v_2,l_2,v_5,l_1,v_1,l_2,v_3,l_1,v_2)$, where at the edge $(v_5, l_1,v_1)$ the KBS is transformed from the kernel-search phase to the kernel-BFS phase that is then guided by $(l_2,l_1)^+$.
When $v_2$ visits itself for the first time, the KBS tries to create index entry $(v_2,(l_2,l_1))\in \lin(v_2)$.
However, this index entry can be pruned by PR1 because of $(v_1,(l_2,l_1))\in \lout(v_2)$ and $(v_1,(l_2,l_1)) \in(v_2)$. 
Then, PR3 is triggered, which means the kernel-BFS from $v_2$ with the kernel $(l_2,l_1)^+$ can terminate.
\end{example}

The indexing algorithm is presented in Algorithm \ref{algo: indexing algorithm}. For ease of presentation, each procedure focuses on the backward case, and the forward case can be obtained by trivial modifications, \textit{e.g.}, replacing in-coming edges with out-going edges. We use the KMP algorithm \cite{Knuth1977FastPM} to compute the minimum repeat of a label sequence, \textit{i.e.}, \texttt{MR()} at line 13 in Algorithm \ref{algo: indexing algorithm}. The indexing algorithm performs backward and forward KBS from each vertex. The KBS from a vertex $v$ consists of two phases: kernel-search (line 6 to line 18) and kernel-BFS (line 25 to line 38).
The kernel-search returns for each vertex $v$ all kernel candidates and a set of frontier vertices $vSet$. 
The kernel-BFS is performed for each kernel candidate, \textit{i.e.}, a BFS with vertices in $vSet$ as frontier vertices guided by a kernel candidate.
PR1 and PR2 are included at line 20, which can be triggered by both kernel-search and kernel-BFS. 
However, RP3 can only be triggered by kernel-BFS, which is implemented at line 36, \textit{i.e.}, if the \texttt{insert} function returns true, then vertices can be pruned during the search. However, PR3 is not related to kernel-search that focuses on generating kernels, such that the result returned by the \texttt{insert} function at line 14 is not taken into account.

\subsection{Analysis of \myIndex\ Index}\label{subsection: analysis}
We present in Theorem \ref{theorem: condensed myIndex} that pruning rules can guarantee the condensed property of the \myIndex\ index, and in Theorem \ref{theorem: correctness of myIndex} that the \myIndex\ index constructed by Algorithm \ref{algo: indexing algorithm} is correct, \textit{i.e.}, sound and complete.
The proofs of Theorem \ref{theorem: condensed myIndex} and \ref{theorem: correctness of myIndex}  are included in our technical report \cite{2203.08606} due to the space limit.

\begin{theorem}\label{theorem: condensed myIndex}
With pruning rules, the \myIndex\ index is condensed.
\end{theorem}

\begin{theorem}\label{theorem: correctness of myIndex}
    Given an edge-labeled graph $G$ and the \myIndex\ index of $G$ with a positive integer $k$ built by Algorithm \ref{algo: indexing algorithm}, there exists a path from vertex $s$ to vertex $t$ in $G$ which satisfies a label constraint $L^+, |L|\leq k$, if and only if one of the following condition is satisfied
    \begin{itemize}
        \item[(1)] $\exists (x,L)\in \lout(s)$ and  $\exists (x,L)\in \lin(t)$;
        \item[(2)] $\exists (t,L)\in \lout(s)$, or $\exists (s,L)\in \lin(t)$. 
    \end{itemize}
\end{theorem}

We analyze the complexity of the \myIndex\ index below.

\textit{Query time.}
Given a query $Q(s,t,L^+)$, the time complexity of Algorithm \ref{algo: query algorithm} is $O(|\lout(s)|+|\lin(t)|)$, because we only need to take $O(|\lout(s)|+|\lin(t)|)$ time to apply the merge join to find $(x,L)\in \lout(s)$ and $(x,L)\in \lin(t)$. Note that index entries in $\lout(s)$ and $\lin(t)$ have already been sorted according to the access id of vertices, such that we  do not need to sort index entries when applying the merge join.

\textit{Index size.}
The index size can be $O(|V|^2 |\Lab|^k)$ in the worst case, since each $\lin(v)$ or $\lout(v)$ can contain $O(|V|C)$ index entries, where $C=O(|\Lab|^k)$ is the number of distinct minimum repeats for all label sequences derived from $|\Lab|$ of length up to $k$. $C$ can be computed as follows, $C = \sum_{i=1}^{k}F(i)$, where $F(i)=|\Lab|^i-(\sum_{j\in f(i), j\neq i} F(j))$ with $F(1)=|\Lab|$ and $f(i)$ the set of factors of integer $i$.

\textit{Indexing time.} 
In Algorithm \ref{algo: indexing algorithm}, we perform a KBS from each vertex, and each KBS consists of two phases: the kernel-search phase and the kernel-BFS phase.
Performing a kernel-search of depth $k$ from a vertex requires $O({|\Lab|}^k{|V|}^k)$ time, and generates $O(|L|^k)$ kernel candidates as discussed in the index size analysis.
Each kernel candidate requires a kernel-BFS taking $O(|E| k)$ time.
Hence the time complexity for performing a KBS is 
$O({|\Lab|}^k{|V|}^k +|L|^k |E| k)$.
The total index time is $O(|V|^{k+1}|\Lab|^k +|L|^k |V||E|k)$
 in the worst case.

Notice that these complexities resemble the complexity classes of previous label-constraint reachability indexes \cite{10.1145/3035918.3035955,10.14778/3380750.3380753} for LCR queries.  

%% file: sections/Experimental_Results.tex
\begin{table}
    \caption{Overview of real-world graphs.}
    \label{table: overview of real-world graphs}
    \resizebox{\linewidth}{!}{
        \begin{tabular}{|c|c|c|c|c|c|c|}
            \hline
            \textbf{Dataset}            & $\bm{|V|}$        & $\bm{|E|}$        & $\bm{|\Lab|}$ &   \begin{tabular}{@{}c@{}} \textbf{Synthetic} \\  \textbf{Labels} \end{tabular} & \begin{tabular}{@{}c@{}} \textbf{Loop} \\ \textbf{Count} \end{tabular}  & \begin{tabular}{@{}c@{}} \textbf{Triangle} \\                                                                  \textbf{Count} \end{tabular}   \\ \hline
            \href{http://konect.cc/networks/advogato/}{Advogato (AD)}                        & 6K                & 51K               & 3             &                                   & 4K            & 98K    \\ \hline
            \href{http://konect.cc/networks/soc-Epinions1/}{Soc-Epinions (EP)}               & 75K               & 508K              & 8             & $\surd$                           & 0             & 1.6M   \\ \hline
            \href{http://konect.cc/networks/munmun_twitter_social/}{Twitter-ICWSM (TW)}      & 465K              & 834K              & 8             & $\surd$                           & 0             & 38K    \\ \hline
            \href{https://snap.stanford.edu/data/web-NotreDame.html}{Web-NotreDame (WN)}      & 325K              & 1.4M              & 8             & $\surd$                           & 27K           & 8.9M   \\ \hline
            \href{https://snap.stanford.edu/data/web-Stanford.html}{Web-Stanford (WS)}       & 281K              & 2M                & 8             & $\surd$                           & 0             & 11M    \\ \hline
            \href{https://snap.stanford.edu/data/web-Google.html}{Web-Google (WG)}           & 875K              & 5M                & 8             & $\surd$                           & 0             & 13M    \\ \hline
            \href{https://snap.stanford.edu/data/wiki-Talk.html}{Wiki-Talk (WT)}             & 2.3M              & 5M                & 8             & $\surd$                           & 0             & 9M     \\ \hline
            \href{https://snap.stanford.edu/data/web-BerkStan.html}{Web-BerkStan (WB)}       & 685K              & 7M                & 8             & $\surd$                           & 0             & 64M    \\ \hline
            \href{https://snap.stanford.edu/data/wiki-topcats.html}{Wiki-hyperlink (WH)}     & 1.7M              & 28.5M             & 8             & $\surd$                           & 4K             & 52M    \\ \hline
            \href{https://snap.stanford.edu/data/soc-Pokec.html}{Pokec (PR)}                 & 1.6M              & 30.6M             & 8             & $\surd$                           & 0             & 32M    \\ \hline
            \href{https://snap.stanford.edu/data/sx-stackoverflow.html}{StackOverflow (SO)}  & 2.6M              & 63.4M             & 3             &                                   & 15M           & 114M   \\ \hline
            \href{https://snap.stanford.edu/data/soc-LiveJournal1.html}{LiveJournal (LJ)}    & 4.8M              & 68.9M             & 50            & $\surd$                           & 0             & 285M   \\ \hline
            \href{http://konect.cc/networks/wikipedia_link_fr/}{Wiki-link-fr (WF)}           & 3.3M              & 123.7M            & 25            & $\surd$                           & 19K           & 30B     \\ \hline
        \end{tabular}
        }
        \vspace*{-0.5cm}
\end{table}

\section{Experimental Evaluation}\label{section: experimental results}
In this section, we study the performance of the \myIndex\  index.
We first used real-world graphs to evaluate the indexing time, index size, and query time of the \myIndex\ index, where we focus on practical graphs and workloads.
Then, we used synthetic graphs to conduct a comprehensive study of the impact of different characteristics on the \myIndex\ index, including label set size, average degree, and the number of vertices.
Finally, we compared the query time of the \myIndex\ index with existing systems, where we additionally consider more types of reachability queries from real-world query logs in order to demonstrate the generality of our approach.

\paragraph{Baselines}
To the best of our knowledge, the \myIndex\ index is the first indexing technique designed for processing recursive label-concatenated reachability queries, and indices for other types of reachability queries are not usable in our context because of specific path constraints defined in the \myQuery\ queries.
Thus, the chosen baselines for the \myIndex\ index are online graph traversals guided by NFAs (see Section \ref{definition: my query}). We consider both BFS (breadth-first search) and BiBFS (bidirectional BFS) as the underlying online traversal methods, and simply refer to the baselines as BFS and BiBFS in this section. We note that DFS (depth-first search) is an alternative to BFS with the same time complexity but is not as efficient as BiBFS.
In addition, we also include an extended transitive closure as a baseline, referred to as ETC. 
The indexing algorithm of ETC performs a forward KBS from each vertex without pruning rules, and records for every reachable pair of vertices $(u,v)$ any k-MR of any path $p(u,v)$.
In ETC, we use a hashmap to store reachable pairs of vertices and the corresponding set of k-MRs.
There are two differences between the indexing algorithm of ETC and the one of the \myIndex\ index: (1) only forward KBS is used for building ETC, instead of forward and backward KBS for the \myIndex\ index, and (2) none of the pruning rules is applied for building ETC.

\begin{figure*}
        \centering
    \includegraphics[width=0.3\textwidth]{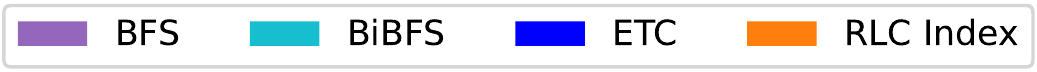}\\
    \begin{minipage}{0.45\textwidth}
        \includegraphics[width=\linewidth]{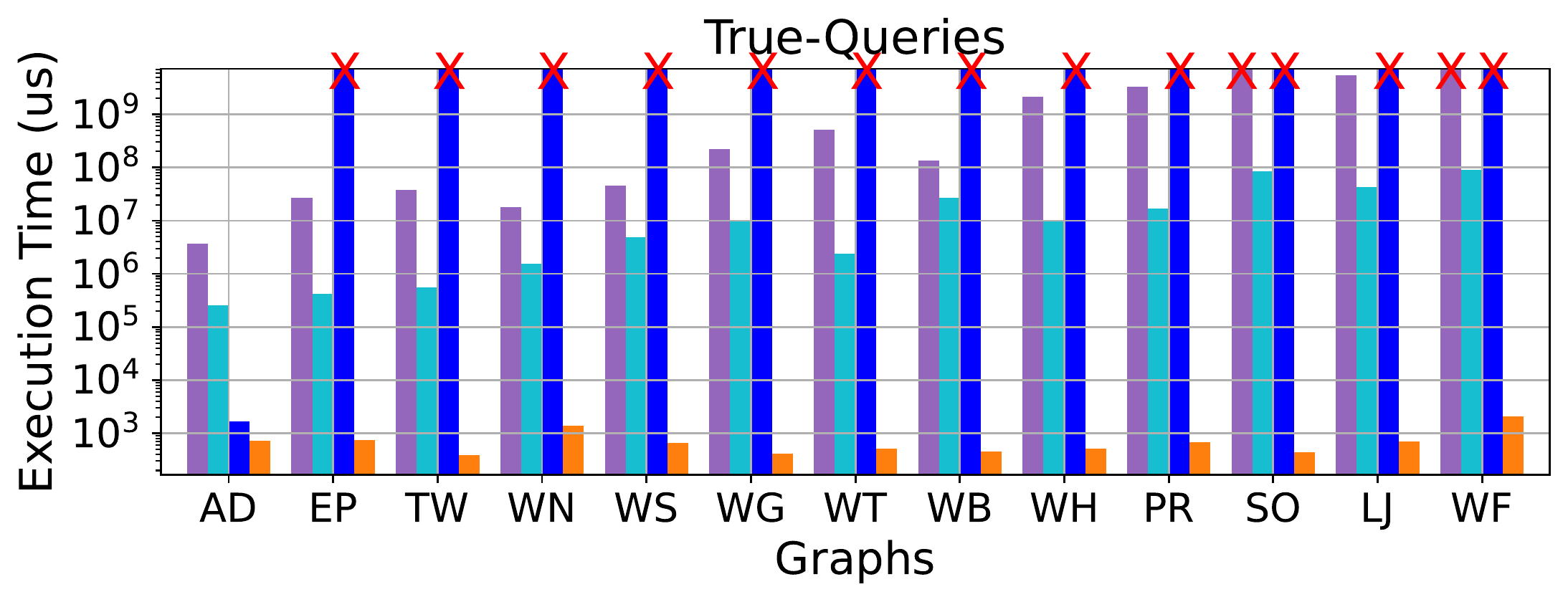}
	\end{minipage}
	\hfill
    \begin{minipage}{0.45\textwidth}
        \includegraphics[width=\linewidth]{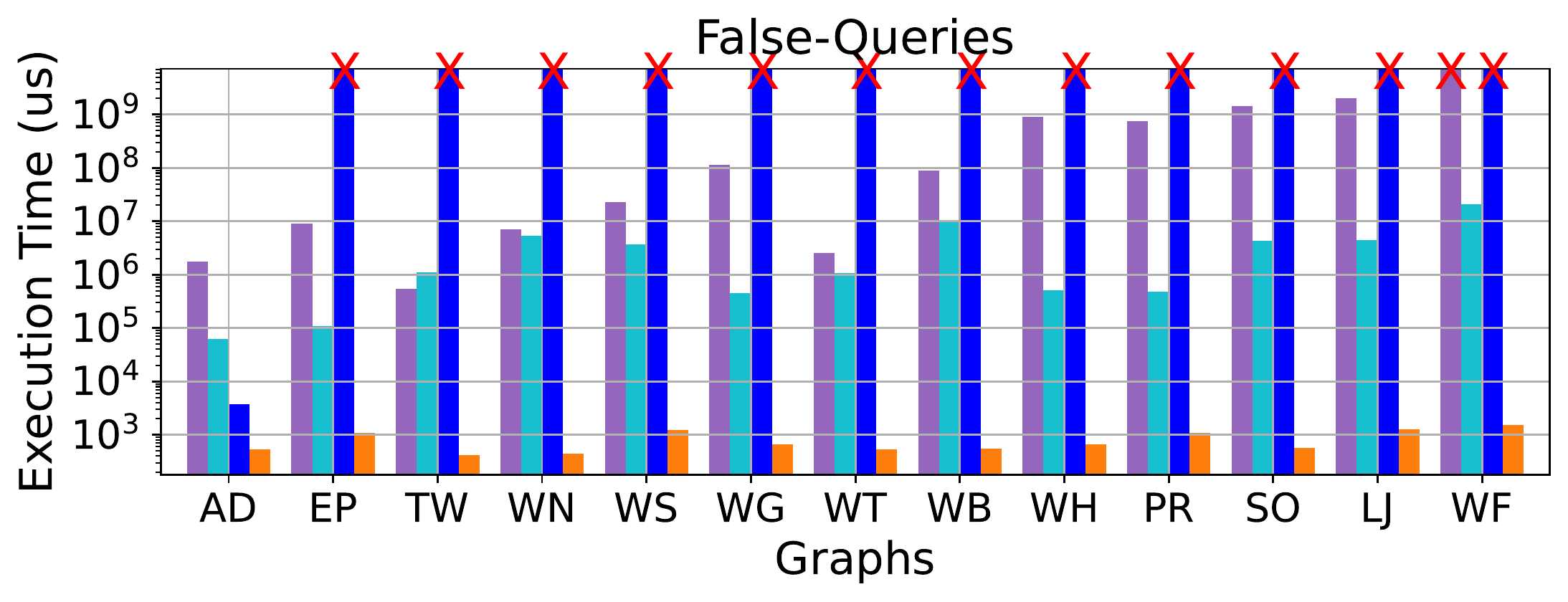}
	\end{minipage}
    \caption{Execution time of 1000 true-queries or 1000 false-queries on real-world graphs.}
    \label{fig: query time results in exp1}      
    \vspace*{-0.5cm}
\end{figure*}

\paragraph{Datasets}
We use real-world datasets and synthetic graphs in our experiments.
We present the statistics of real-world datasets in Table \ref{table: overview of real-world graphs} (sorted according to $|E|$), which are from either the SNAP \cite{snapnets} or the KONECT \cite{10.1145/2487788.2488173} project.
We also include for each graph the loop count (cycles of length $1$) and the triangle count (cycles of length $3$) shown in the last two columns in Table \ref{table: overview of real-world graphs}.
We generate synthetic labels for graphs that do not have labels on their edges, indicated by the column of synthetic labels in Table \ref{table: overview of real-world graphs}.
The edge labels have been generated according to the Zipfian distribution \cite{BBCFLA17} with exponent $2$.
The synthetic graphs used in our experiments follows two different modes, namely the \textit{Erdős–Rényi} (ER) model and the \textit{Barabási-Albert} (BA) model.
We use JGraphT \cite{10.1145/3381449} to generate the ER- and BA-graphs.
The main difference between ER-graphs and BA-graphs is that ER-graphs have an almost uniform degree-distribution while BA-graphs have a skew in it because BA-graphs contain a complete sub-graphs.
The method to assign labels to edges in synthetic graphs is the same as the one used for real-world graphs.

\paragraph{Query generation}
As a common practice to evaluate a reachability index in the literature, \textit{e.g.}, \cite{10.1145/3035918.3035955,10.14778/3380750.3380753,10.1145/3451159}, we generate for each real-word graph comprehensive query sets, each of which contains 1000 true-queries and 1000 false-queries, respectively. 
We explain the method for query generation as follows.
We uniformly select a source vertex $s$ and a target vertex $t$, and also uniformly choose a label constraint $L^+$. 
Then, a bidirectional breadth-first search is conducted to test whether $s$ reaches $t$ under the constraint of $L^+$. 
If the test returns $true$, we add $(s,t,L^+,true)$ to the true-query set, otherwise we add it into the false-query set. 
After that, we generate another $(s,t,L^+)$, and repeat the above procedure until the completion of the two query sets.

\paragraph{Implementation and Setting}
Our open-source implementation has been done in Java 11, spanning baseline solutions and the \myIndex\ index.
We run experiments on a machine with 8 virtual CPUs of 2.40GHz, and 128GB main memory, where the heap size of JVM is configured to be 120GB.

\subsection{Performance on Real-World Graphs}\label{subsection: experiments with real-world graphs}
In this section, we analyze the performance of the \myIndex\ index on real-world graphs.
We first compare the \myIndex\ index with ETC in terms of indexing time and index size, and with BFS and BiBFS in terms of query time. 
The recursive $k$ value is set to $2$, and each query in the respective workload has a recursive concatenation of $2$ labels. The goal of the first experiment is to understand the performance of the \myIndex\ index for practical \myQuery\ query workloads whose  length of recursive concatenation is at most $2$, as in practical property paths \cite{10.1145/3308558.3313472}.
After that, we conduct experiments on real-world graphs with different recursive $k$ values to understand the impact for general queries.

\begin{table}
        \centering
        \caption{Indexing time (IT) and index size (IS).} \label{table: index time and index size}
        \resizebox{0.7\linewidth}{!}{
        \begin{tabular}{|c|rr|rr|}
            \hline
            \textbf{Dataset}      & \multicolumn{2}{c|}{\textbf{ \myIndex\ Index}} & \multicolumn{2}{c|}{\textbf{ETC}} \\
                                 & IT (s)                   & IS (MB)                & IT (s)         & IS (MB)         \\\hline
                AD               & 0.7                 & 1.9               &  2216.1     & 2798.7         \\\hline
                EP               & 22.6                & 29.3              &  -          &  -          \\\hline
                TW               & 8.1                 & 93.5              &  -          &  -          \\\hline
                WN               & 33.1                & 122.6             &  -          &  -          \\\hline
                WS               & 53.5                & 173.9             &  -          & -           \\\hline
                WG               & 101.3               & 403.6             &  -          & -           \\\hline
                WT               & 812.9               & 607.1             &  -          & -           \\\hline
                WB               & 167.1               & 474.2             &  -          & -          \\\hline
                WH               & 3707.2              & 1319.1            &  -          & -          \\\hline
                PR               & 3104.1              & 1212.6            &  -          & -           \\\hline
                SO               & 57072.5             & 844.2             &  -          & -           \\\hline
                LJ               & 18240.9             & 6248.1            &  -          & -           \\\hline
                WF               & 51338.7             & 6467.9            &  -          & -          \\\hline
        \end{tabular}}    
            \vspace*{-0.5cm}
\end{table}

\textit{Indexing time.}
Table \ref{table: index time and index size} shows the indexing time of the \myIndex\ index and ETC on real-world graphs, where ``-" indicates that the method timed out on the graph.
Building ETC cannot be completed in 24 hours for real-world graphs (or it runs out of memory) except for the AD graph with the least number of edges.
The \myIndex\ index for the AD graph can be built in 0.7s, leading to a four-orders-of-magnitude improvement over ETC.
The indexing time improvement of the \myIndex\ index over ETC is mainly due to the pruning rules that skip vertices in graph traversals when building the \myIndex\ index. Thus, this experiment also shows the significant impact of pruning rules in terms of indexing time.
The indexing time of the \myIndex\ index for the first 10 graphs is at most 1 hour. The last three graphs, \textit{i.e.}, the SO graph, the LJ graph, and the WF graph are more challenging than the others, not only because they have more vertices and edges, but also because they have a larger number of loops and triangles, as shown in Table \ref{table: overview of real-world graphs}.
The SO graph has the longest indexing time due to its highly dense and cyclic character, \textit{i.e.}, it has 15M loops and 114M triangles. 
Although the WF graph has much fewer loops than the SO graph, it contains 30B triangles.
Consequently, the indexing time of the WF graph is at the same order of magnitude as the one of the SO graph.
While it has more vertices, triangles, and edge labels than the SO graph, the LJ graph requires a lower indexing time.

\textit{Index size.}
Table \ref{table: index time and index size} shows the size of the \myIndex\ index for real-world graphs.
The size of the \myIndex\ index is much smaller than the size of ETC for the AD graph (that is the only graph ETC can be built within 24 hours).
The index size improvement is because of not only the indexing schema of the \myIndex\ index but also the application of pruning rules that can avoid recording redundant index entries.
The effectiveness of pruning rules can also be observed between the PR graph and the SO graph. 
Although the SO graph is larger than the PR graph in terms of the number of vertices and the number of edges, the index size of the SO graph is smaller than the index size of the PR graph.
Thus, this experiment also demonstrates the significant impact of pruning rules in terms of index size.

\textit{Query time.}
Fig. \ref{fig: query time results in exp1} shows the execution time of the true-query set and the false-query set.
In general,  the execution time of a query set of 1000 queries using the \myIndex\ index is around 1 millisecond for all the graphs in Table \ref{table: overview of real-world graphs}, except for the WF graph (that has the largest number of edges) for which around 2 milliseconds. 
Query execution using BFS times out for the true-queries on both the SO graph and the WF graph, and for the false queries on the WF graph.
In addition,  we only report the query time of ETC for the AD graph as ETC cannot be built for all the other graphs.
As shown in Fig. \ref{fig: query time results in exp1}, the \myIndex\ index shows an up to six-orders-of-magnitude improvement over BFS and four-orders-of-magnitude improvement over BiBFS.
The query time using the \myIndex\ index is slightly faster than ETC, because the larger number of reachable pairs of vertices recorded in ETC leads to a slight overhead of checking for the reachability between a source $s$ and a target $t$ and also the existence of a minimum repeat of a path from $s$ to $t$.

\begin{figure}
    \centering
    \includegraphics[width=\linewidth]{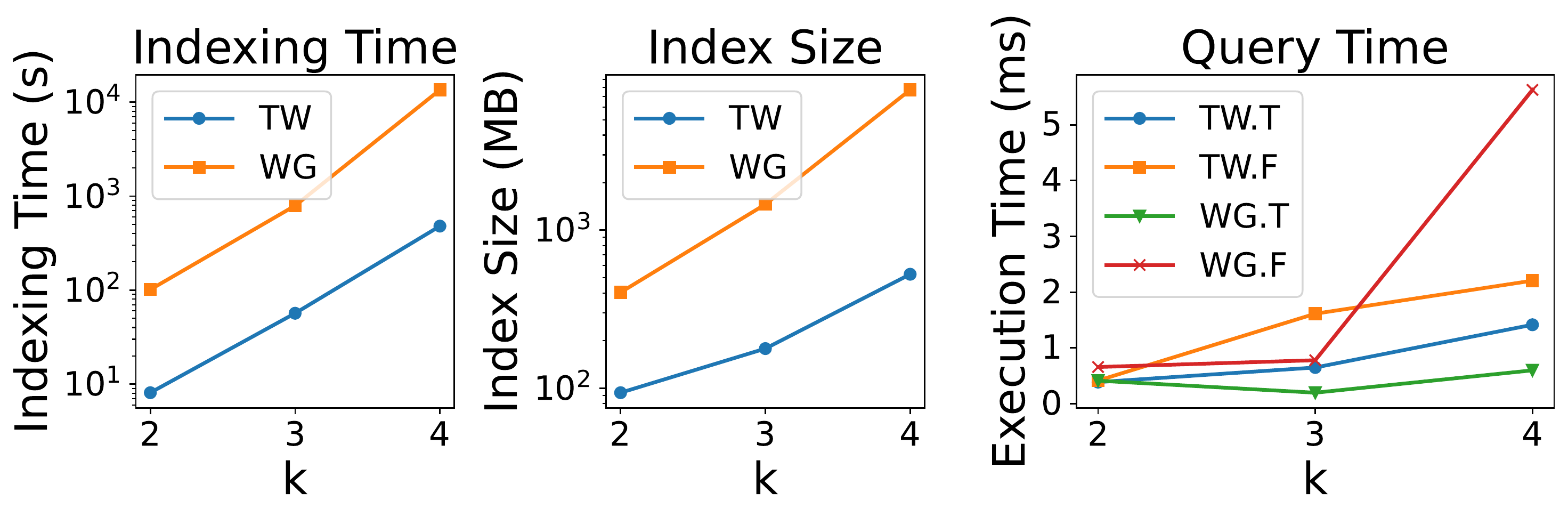}
    \caption{\myIndex\ index performance with different recursive $k$ values.}
    \label{fig:real-world_graphs_various_k}
    \vspace*{-0.5cm}
\end{figure}

\textit{Different recursive k values.}
We analyze the impact of the recursive $k$ values on real-world graphs.
For space reasons, we have focused on the TW and WG datasets, \textit{i.e.}, graphs from Twitter and Google.
The indexing results are shown in Figure \ref{fig:real-world_graphs_various_k}, where we also report the query time of  $1000$ true-queries and $1000$ false-queries with a recursive concatenation of $2$, $3$, and $4$ labels.
As expected, the indexing time and index size of the \myIndex\ index increase when the recursive $k$ value increases, and the larger index size with more path constraints can also lead to the increase of query time. 
The main reason is due to the fact that the number of path-constraints (kernels) will exponentially grow as the increasing of k, and the kernel-search phase of the indexing algorithm needs to take into consideration the potential path-constraints that exist in the graph. 
Notice that the increasing rate of index size is much slower than the increasing rate of indexing time, which means that the number of paths that can satisfy path constraints of recursive concatenation is not increasing as the increase of the length of the constraints, \textit{i.e.}, the recursive $k$ value. 
The main reason is that in real-world graphs only only a few labels have a large number of occurrence (the Zipfian distribution) as shown in existing benchmarks \textit{e.g.}, gMark \cite{BBCFLA17}. 
Consequently, a long concatenation of edge labels cannot repeat frequently due to the lack of desired labels. Thus, \myIndex\ queries with a large recursive k value may not need indexing.

\textit{Summary and outlook}.
Our indexing algorithm designed for an arbitrary recursive $k$ value can efficiently build the \myIndex\ index for processing practical \myQuery\ queries that are difficult to evaluate in modern graph query engines.
Specifically, the recursive concatenation length of \myQuery\ queries in recent open-source query logs \cite{10.1145/3308558.3313472} is not larger than $2$ and such practical \myQuery\ queries appear quite often in time-out logs. 
As shown in Table \ref{table: index time and index size} and Fig. \ref{fig: query time results in exp1}, the \myIndex\ index with a recursive $k$ of $2$ can be efficiently built even for large and highly dense real-world graphs, and can significantly improve the processing time of such timed out queries in practice.

\subsection{Impact of Graph Characteristics}\label{subsection: experiments with synthetic graph}
In this section, we focus on analyzing the performance of the \myIndex\ index on synthetic graphs (ER-graphs and BA-graphs) with different characteristics, namely average degree, label set size, and the number of vertices.
The recursive $k$ value is set to 2 in this section, and more experiments about different $k$ values on synthetic graphs are presented in our technical report \cite{2203.08606}.
We generate for each graph a query set of 1000 true-queries and a query set of 1000 false-queries,  which are referred to as $ER.T$ and $ER.F$ for an ER-graph, and $BA.T$ and $BA.F$ for a BA-graph.

\subsubsection{Impact of label set size and average degree}
\begin{figure*}
    \begin{minipage}{\textwidth}
        \centering
        \includegraphics[width=0.3\linewidth]{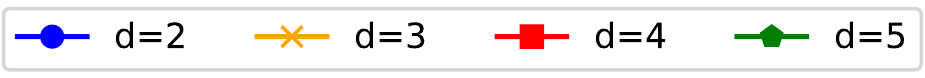}\\
        \centering
        \begin{minipage}{0.2\linewidth}
            \includegraphics[width=\linewidth]{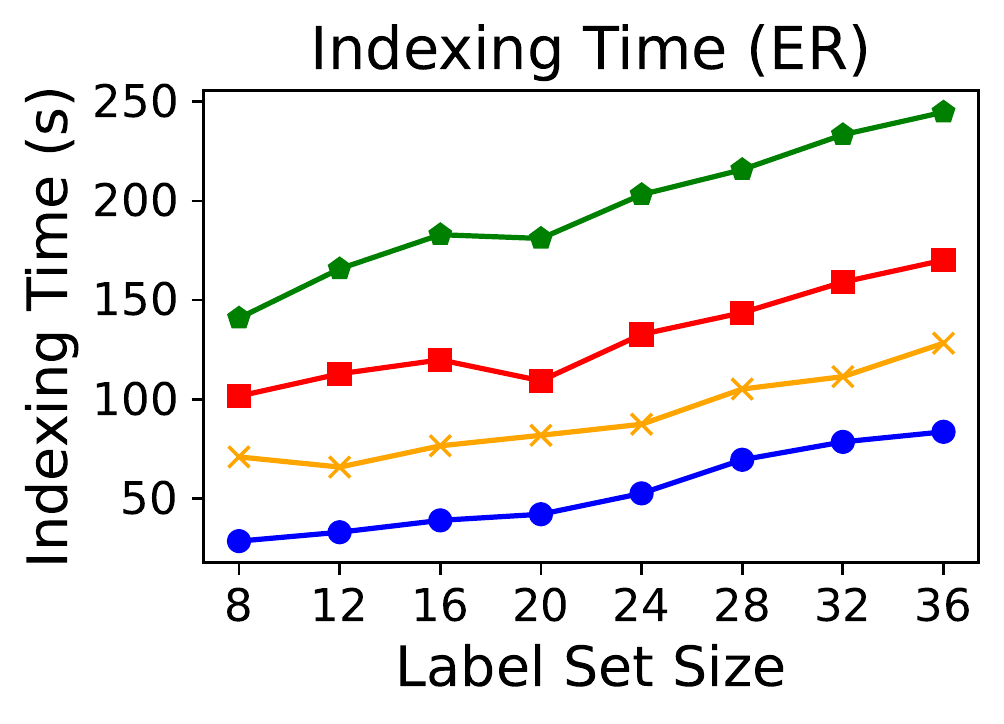}
    	\end{minipage}
    	\hfill
    	\begin{minipage}{0.2\linewidth}
            \includegraphics[width=\linewidth]{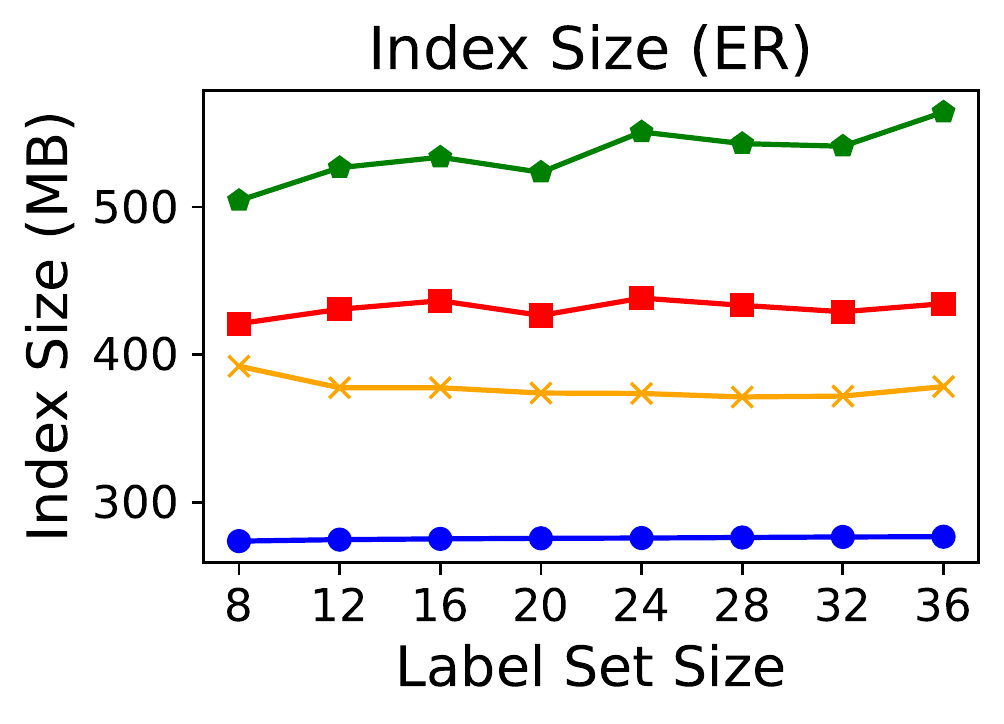}
    	\end{minipage}
    	\hfill
    	\begin{minipage}{0.2\linewidth}
            \includegraphics[width=\linewidth]{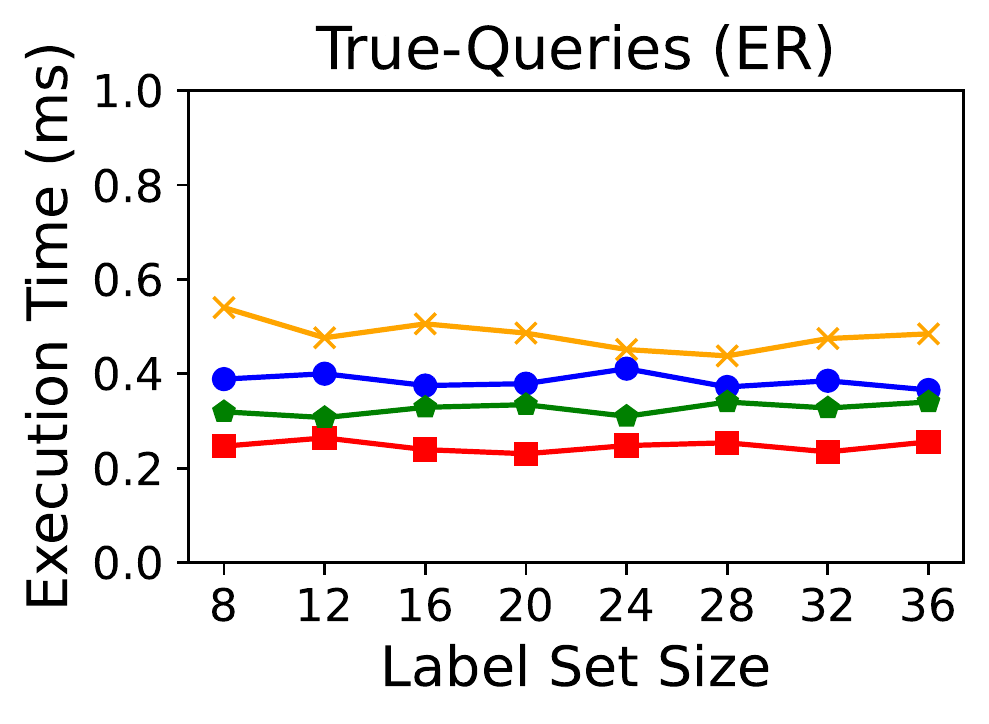}
    	\end{minipage}
    	\hfill
    	\begin{minipage}{0.2\linewidth}
            \includegraphics[width=\linewidth]{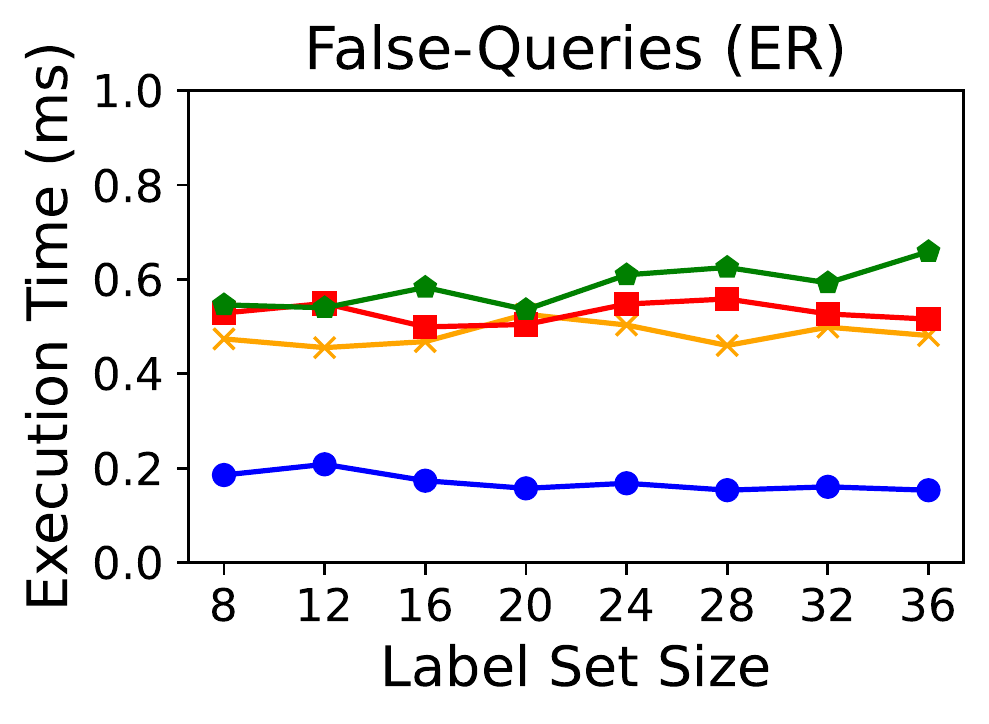}
    	\end{minipage}
        \centering
        \begin{minipage}{0.2\linewidth}
            \includegraphics[width=\linewidth]{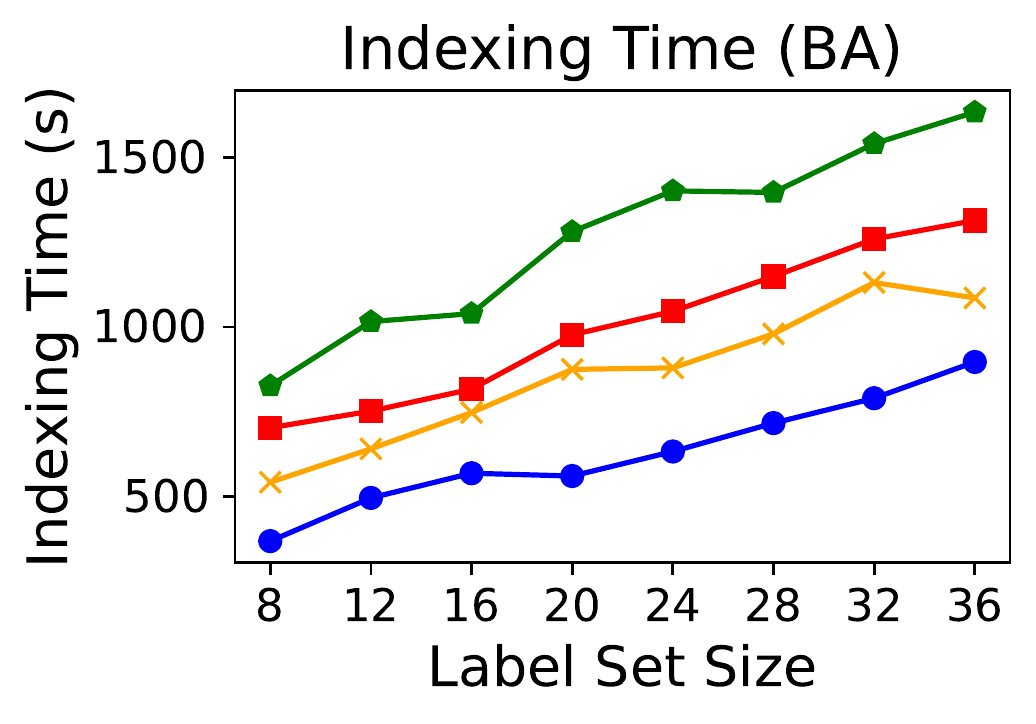}
    	\end{minipage}
    	\hfill
        \begin{minipage}{0.2\linewidth}
            \includegraphics[width=\linewidth]{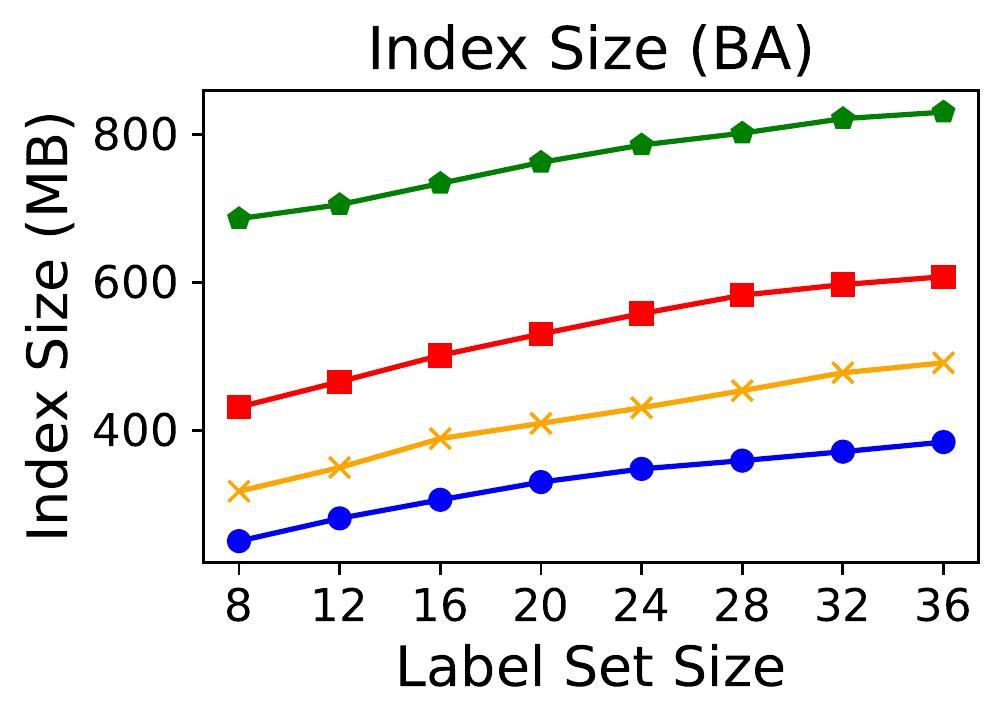}
    	\end{minipage}
    	\hfill
    	\begin{minipage}{0.2\linewidth}
            \includegraphics[width=\linewidth]{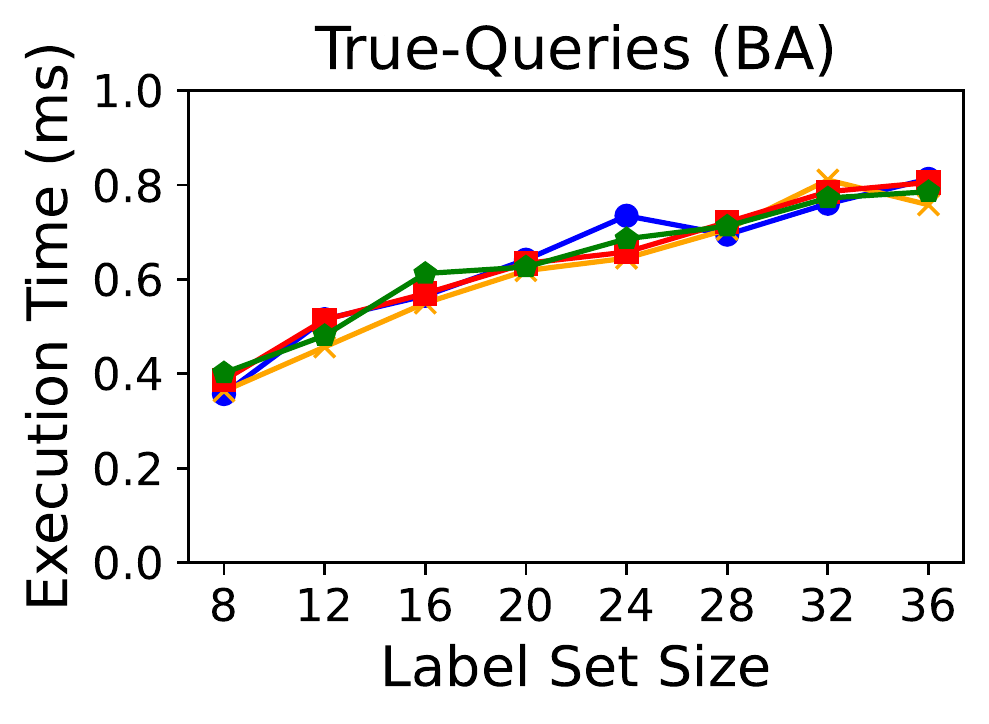}
    	\end{minipage}
    	\hfill
    	\begin{minipage}{0.2\linewidth}
            \includegraphics[width=\linewidth]{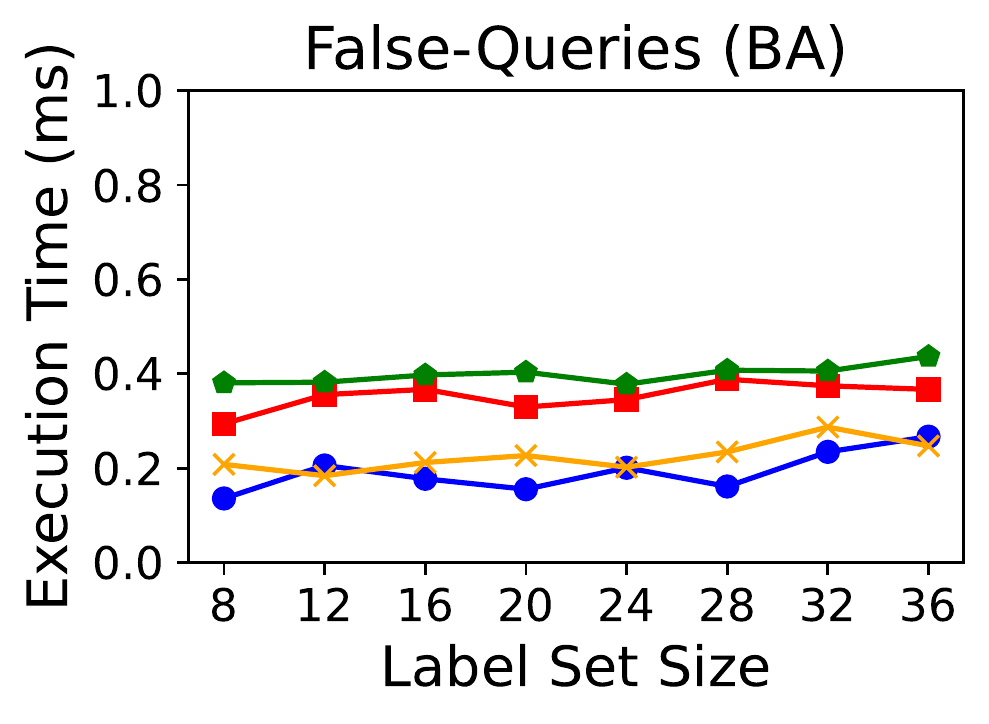}
    	\end{minipage}
        \caption{Indexing time, index size, and execution time for graphs with $|V|=$ 1M, varying $d$, and varying $|\Lab|$.}
        \label{fig: exp3}          
        \vspace*{-0.5cm}
    \end{minipage}
\end{figure*}

In this experiment, we use BA-graphs and ER-graphs with 1M vertices, and we vary the average degree $d$ in $(2,3,4,5)$, and label set size $|\Lab|$ in $(8,12,16,20,24,28,32,36)$, \textit{e.g.}, a graph with $d=5$ and $|\Lab| = 16$ has 1M vertices, 5M edges, and 16 distinct edge labels.
We aim at analyzing indexing time, index size, and query time of the \myIndex\ index as the increase of $d$  and $|\Lab|$. 
The experimental results for ER-graphs and BA-graphs are reported in Fig. \ref{fig: exp3}.
We discuss the results below.

\begin{figure*}
    \centering
    \begin{minipage}{0.20\textwidth}
        \includegraphics[width=\linewidth]{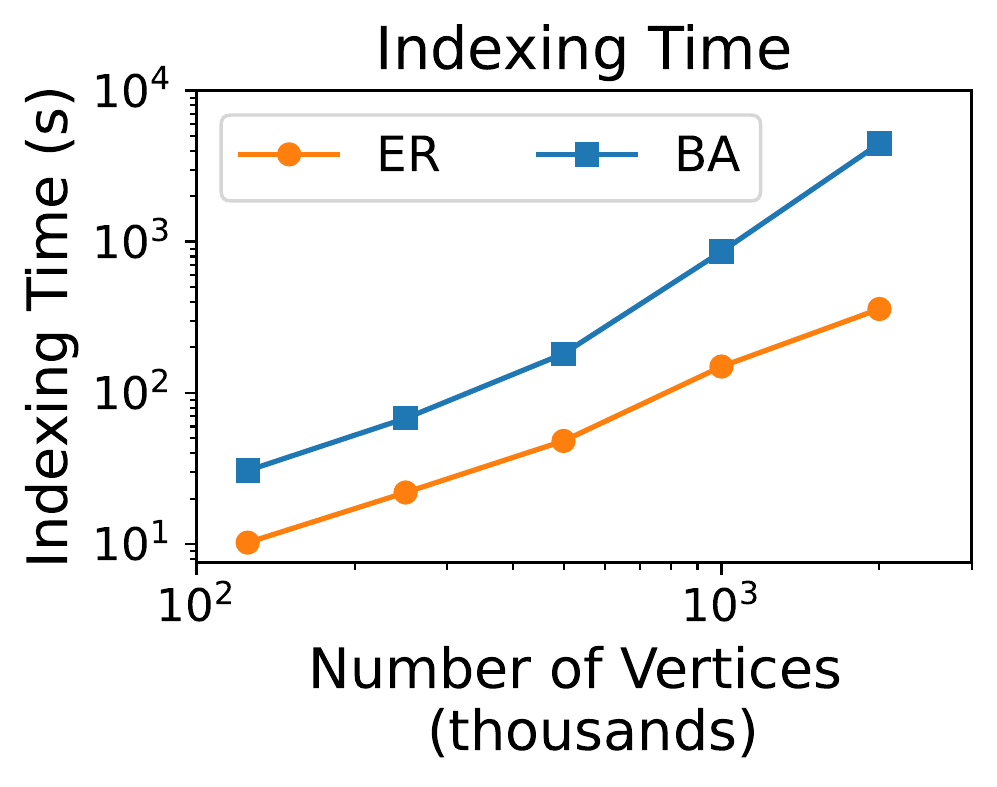}
	\end{minipage}
	\hfill
    \begin{minipage}{0.20\textwidth}
        \includegraphics[width=\linewidth]{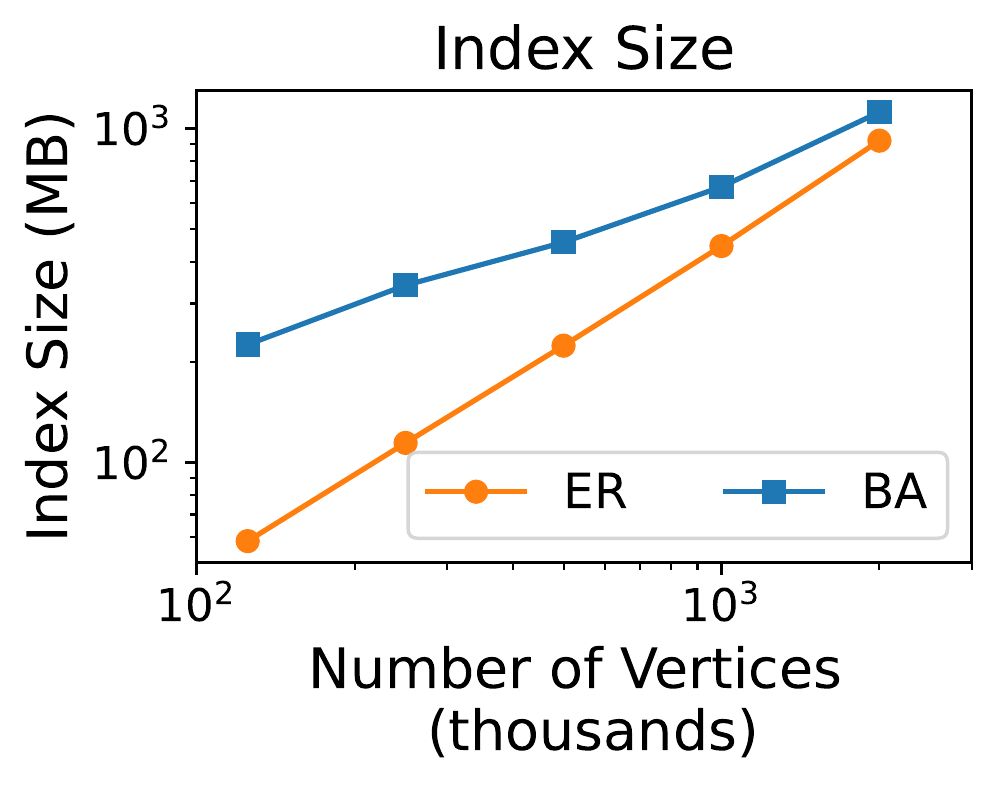}
	\end{minipage}
	\hfill
    \begin{minipage}{0.20\textwidth}
        \includegraphics[width=\linewidth]{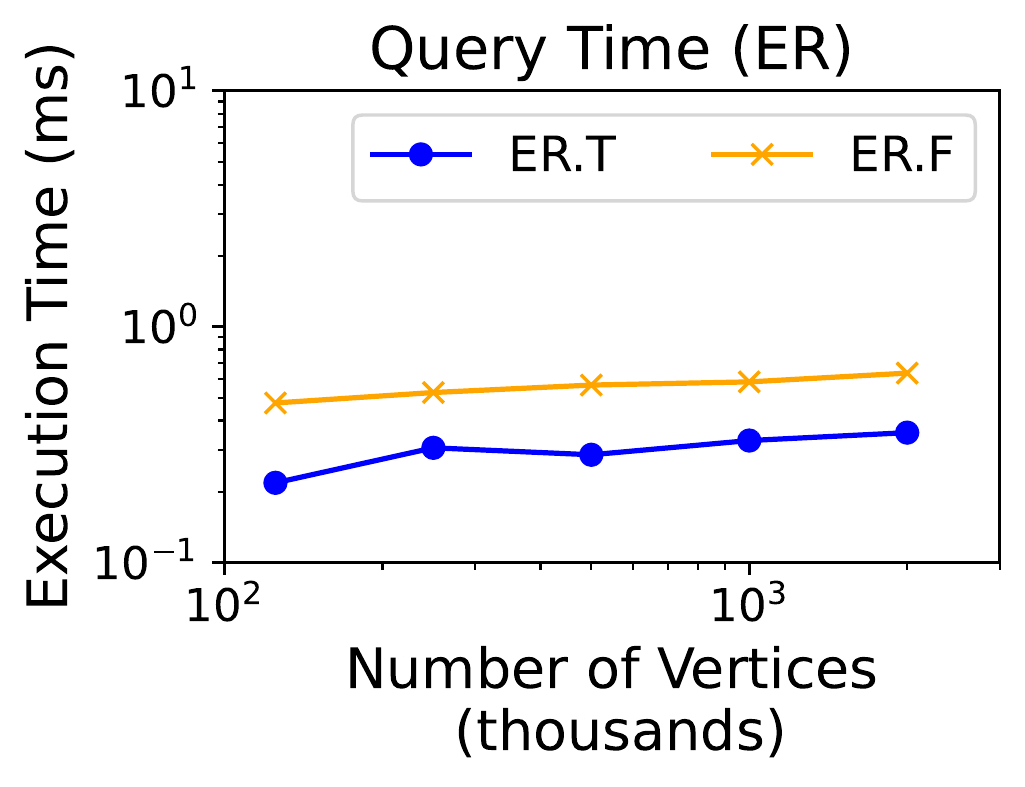}
	\end{minipage}
	\hfill
	\begin{minipage}{0.20\textwidth}
        \includegraphics[width=\linewidth]{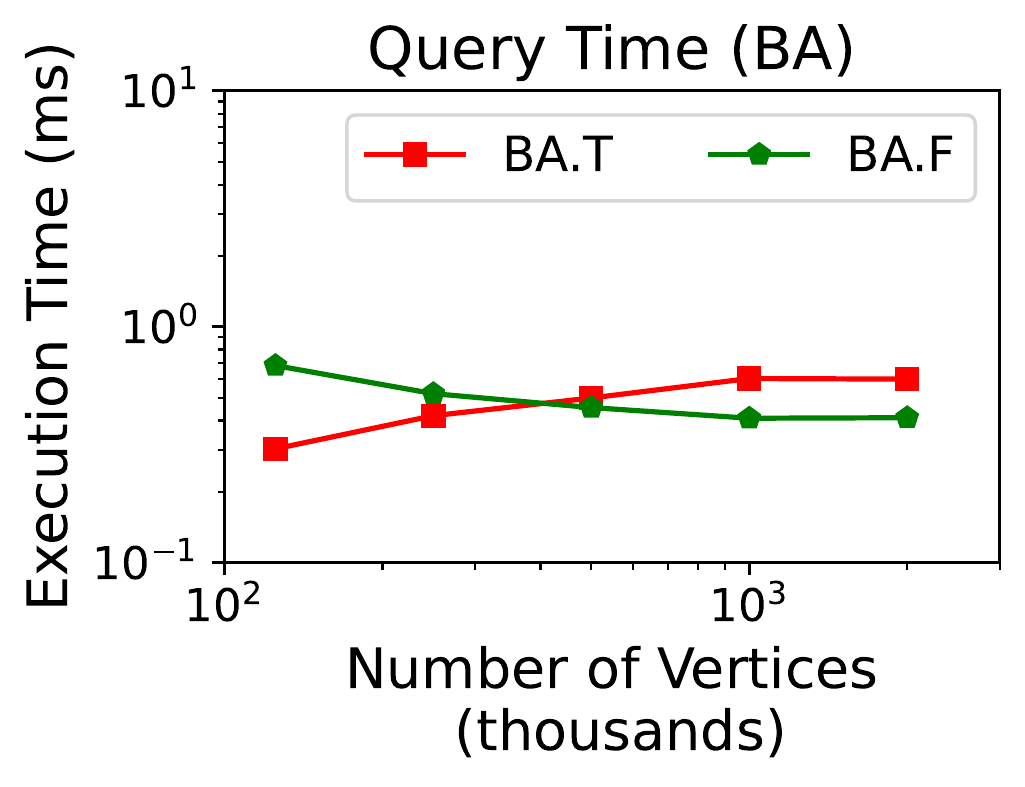}
	\end{minipage}
    \caption{Indexing time, index size, and query execution time for graphs with $d=5$, $|\Lab|=16$, and varying $|V|$.}\label{fig: exp4 of scalability}
    \vspace*{-0.5cm}
\end{figure*}

\textit{Indexing time.}
We observe in Fig. \ref{fig: exp3} that the indexing time for the used ER-graphs and BA-graphs with a fixed $d$  shows a linear increase (for most cases)  as $|\Lab|$ increases.
This can be understood as follows.
When $|\Lab|$ increases, the number of possible minimum repeats increases, requiring more time for the kernel-search phase of a KBS in the indexing algorithm to traverse the graph and generate potential kernel candidates, resulting in more kernel-BFS executions. 
The total number of possible minimum repeats can be  quadratic in $|\Lab|$ in the worst case when the input parameter $k$ of the \myIndex\ index is two. 
Furthermore, because there are more edges to traverse, the indexing time for both ER-graphs and BA-graphs with a fixed $|\Lab|$ increases linearly as $d$ increases.

\textit{Index size.}
As illustrated in Fig. \ref{fig: exp3}, an increase in average degree $d$ can result in a larger index size for both ER-graphs and BA-graphs.
The reason is that a vertex $s$ can reach a vertex $t$ through more paths, leading to a higher number of minimum repeats being recorded.
As the size of the label set grows, the corresponding impact on ER-graphs is different from the one on BA-graphs. Specifically, the increase is negligible for ER-graphs with a small $d$, \textit{e.g.}, $2$, and becomes more noticeable for ER-graphs with a large $d$, \textit{e.g.}, $5$. 
For any $d$, however, we see a clear linear increase in index size with the growth in $|\Lab|$ for BA-graphs.
This is because a BA-graph comprises a complete sub-graph, and vertices inside the complete sub-graph have higher degrees. Therefore, the KBSs executed from such vertices can create more index entries as $|\Lab|$ grows, as it can reach other vertices through paths with more distinct minimum repeats.
However, because of the uniform degree distribution, the increase in the number of minimum repeats of paths from a vertex $s$ to a vertex $t$ due to an increase in $|\Lab|$ is not significant for ER-graphs when $d$ is small, but the corresponding impact becomes stronger when $d$ is larger.

\textit{Query time.}
As shown in Fig. \ref{fig: exp3} the growth of $|\Lab|$ has a different impact on query time.
More precisely, when $|\Lab|$ rises, the execution time of both true- and false-queries for ER-graphs remains steady.
When it comes to BA-graphs, increasing $|\Lab|$ can lead to a minor boost in true-query  execution time but has almost no impact on false query execution time.
This is due to the fact that the vertices in the complete sub-graph of a BA-graph can reach much more vertices than the vertices outside the complete sub-graph in the BA-graph, leading to a skew in the distribution of vertices in index entries, \textit{i.e.}, many index entries have the same vertex.
Furthermore, when $|\Lab|$ grows, the number of minimum repeats also increases, which makes the skew higher.
As a result, processing true-queries will encounter situations where the query algorithm searches for a particular minimum repeat in a significant number of index entries with the same vertex.
However, for false-queries, the query result can be returned instantly if there are no index entries with the same vertex.
Fig. \ref{fig: exp3} also shows that $d$ has a negligible impact on the execution time of the true-queries on the BA graph.
The main reason is that the number of index entries for some vertices in the BA graph does not increase significantly w.r.t the increase of $d$ because of the skew in the degree distribution and a fixed $|\Lab|$.

\subsubsection{Scalability}
In this experiment, we use BA-graphs and ER-graphs with average degree $5$, $16$ edge labels, and vary the number of vertices in (125K, 250K, 500K, 1M, 2M). 
The goal is to analyze the scalability of the \myIndex\ index in terms of $|V|$.
The results of indexing time, index size, and query time for both ER-graphs and BA-graphs are reported in Fig. \ref{fig: exp4 of scalability}.

\textit{Indexing time and index size.}
Fig. \ref{fig: exp4 of scalability} shows that indexing time and index size grows  with the increase of the number of vertices.
The main reason is that graphs with more vertices require more KBS iterations, which increases indexing time and also the number of index entries. 
We observe that indexing BA-graphs is more expensive than indexing ER-graphs because of the presence of a complete sub-graph in the former.
We also observe that the different topological structures between BA-graphs and ER-graphs have different impacts on the increasing rate of index size w.r.t $|V|$. The uniform degree distribution in ER-graphs makes the index size increase with a sharper rate than the one of BA-graphs, because BA-graphs contain a significant number of vertices of high degree, which is also growing as $|V|$ increases, and the indexing algorithm starts building the index from these vertices such that index entries containing these vertices can be leveraged to prune redundant index entries that need to be inserted later on.

\textit{Query time.}
Fig. \ref{fig: exp4 of scalability} shows that query time on the ER-graphs and true-query time on the BA-graphs slightly increase as the number of vertices grows, as expected on larger graphs.
In addition, the false-query time is higher than the true-query time on ER-graphs, and the true-query time is higher than the false-query time on BA-graphs.
We interpret the results as follows.
Given a query $(s,t,L^+)$ on a graph which has a uniform distribution in vertex degree, the index entries $(v,mr)$ in both $\lout(s)$ and $\lin(t)$ also have a uniform distribution in terms of $v$. Thus, the query algorithm (based on merge join) need to perform an exhaustive search in $\lout(s)$ and $\lin(t)$ to find index entries with the same vertex $v$, which results in false-queries taking longer time to execute than true-queries.
However, when the distribution of vertex degree is skewed in BA-graphs, the index entries in $\lout(s)$ and $\lin(t)$ are dominated by vertices of high degree.
In addition, as there are many paths passing through high-degree vertices with distinct minimum repeats, the number of index entries with such vertices is also large.  
Thus, the query algorithm can perform a faster search for false-queries than true-queries, because the number of distinct vertices in both $\lout(s)$ and $\lin(t)$ is not large. 
Notice that the number of vertices of high degree also increases as $|V|$ increases. For false-queries in BA-graphs, the dominance of index entries with high-degree vertices becomes stronger as $|V|$ increases. Consequently, the number of index entries in $\lout(s)$ and $\lin(t)$ with distinct vertex id can decrease, which makes the merge-join algorithm execute faster.
The query algorithm, on the other hand, needs to select a specific $mr$ among index entries with vertex $u$ of a high degree to process true-queries, which takes more time.

\subsection{Comparison with Existing Systems}\label{subsection: speedup on graph engines}
In this section, we focus on evaluating how much improvement the \myIndex\ index can provide over mainstream graph processing systems. We recall that many current graph query engines fail to evaluate \myQuery\ queries, thus we focused on three systems that are able to evaluate these queries on property graphs and RDF graphs. 
In order to show how the index performs with varying types of queries, such as longer concatenations or path queries frequently occurring in real-world query logs, we consider the following queries: \textbf{Q1} being a single label under the Kleene plus $a^+$; \textbf{Q2} consisting of a concatenation of length 2 $(a\circ b)^+$; \textbf{Q3} having the expression $(a\circ b \circ c)^+$ thus a concatenation of length 3. 
In this case, we build the index with $k=3$ to support all the three \myQuery\ queries, especially Q3 having the longest concatenation.
For the sake of completeness, an extended reachability query with the constraint  $a^+\circ b^+$ is also included in this experiment, referred to as \textbf{Q4}.
We have employed Q4 to study the generality and applicability of the \myIndex\ index to a wide range of regular path queries in real-world graph query logs \cite{10.1145/3308558.3313472}. 
To deal with this query, we use the \myIndex\ index in combination with an online traversal to continuously check whether intermediately visited vertices can satisfy the path constraint.

Three graph engines, including commercial and open-sourced ones, used in the experiments have been selected as representatives of the few available graph engines capable of evaluating \myQuery\ queries. 
We do not reveal the identity of all the systems as some are proprietary and we cannot release performance data.
Anonymized engines are referred to as Sys1 and Sys2 in the results, and the third one is Virtuoso Open-Source Edition (v7.2.6.3233).
For the systems that support multi-threaded query evaluation, we set the system to single-threaded to ensure a fair comparison with our approach, which is single-threaded only.
For Virtuoso, we disabled the transaction logging to avoid additional overheads, and configured it to work entirely in memory by setting the maximum amount of memory for transitive queries to the available memory of the server. 
Note that these systems have their own indexes by default, which we leave the configuration unchanged, \textit{e.g.}, Virtuoso 7 has column-wise indexes by default. These indexes are not suitable for  \myQuery\ queries, as confirmed by our comparative analysis.

\begin{table}
\caption{Speed-ups (SU) and workload size break-even points (BEP) of the \myIndex\ index  over graph engines.}\label{table: speed-up}
\resizebox{\linewidth}{!}{
\begin{tabular}{|c|cccccc|cc|}
\hline
\multirow{3}{*}{\textbf{\begin{tabular}[c]{@{}c@{}}\\ Sys.\end{tabular}}} & \multicolumn{6}{c|}{\textbf{RLC Query}}                                                                   & \multicolumn{2}{c|}{\textbf{\begin{tabular}[c]{@{}c@{}}Extended \\ Query\end{tabular}}} \\ \cline{2-9} 
                                                                               & \multicolumn{2}{c|}{\textbf{Q1}}  & \multicolumn{2}{c|}{\textbf{Q2}}   & \multicolumn{2}{c|}{\textbf{Q3}} & \multicolumn{2}{c|}{\textbf{Q4}}                                                        \\
                                                                               & SU   & \multicolumn{1}{c|}{BEP}   & SU    & \multicolumn{1}{c|}{BEP}   & SU              & BEP            & SU                                          & BEP                                       \\ \hline
Sys1                                                                           & 1200x & \multicolumn{1}{c|}{84100} & 10400x & \multicolumn{1}{c|}{34000}   & 18400x           & 9400           & 34000x                                         & 300                                       \\ \hline
Sys2                                                                           & 3000x   & \multicolumn{1}{c|}{34900} & 202000x  & \multicolumn{1}{c|}{1700}  & 1300000x            & 130            & 104000x                                        & 98                                        \\ \hline
Virtuoso                                                                           & 597x  & \multicolumn{1}{c|}{180000}  & 4900x  & \multicolumn{1}{c|}{71700} & 38100000x           & 5              & -                                           & -                                         \\ \hline
\end{tabular}}   
    \vspace*{-0.5cm}
\end{table}

We use the WN graph as a representative of real-world graphs, which has a moderate number of vertices and edges, along with a sufficient number of cycles to evaluate.
We build one \myIndex\ index with the parameter $k=3$ for the WN graph and use it to process all four queries. In this case, the \myIndex\ index of the WN graph can be built in 5.9 minutes and takes up 821 megabytes.
We run each query using each system within the 10-minutes time limit, and we repeat the execution 20 times and report the median of the query execution time.
We use two metrics to evaluate the improvement of the \myIndex\ index, namely speed-ups (SU) and the workload size break-even points (BEP).
SU shows the query time improvement of the \myIndex\ index over an included graph system. 
BEP indicates the number of queries that make the indexing time of the \myIndex\ index pay off.
Table \ref{table: speed-up} shows the results, where "-" indicates that the query execution of this system timed out.
The \myIndex\ index shows that using a single index lookup can gain significant speed-ups over included systems for processing Q3, which has the longest concatenation under the Kleene plus.
We use the BEP value to understand the amortized cost of using the \myIndex\ index to accelerate Q3 processing.
In particular, executing Q3 130 times on Sys2 is equivalent to the time it takes to build and query the \myIndex\ index the same number of times.
The \myIndex\ index can also significantly improve the execution time for Q1, Q2, and Q4.

%% file: sections/Conclusion.tex
\section{Conclusion}\label{section: conclusion}
In this paper, we introduced \myQuery\ queries, reachability queries with a path constraint based on the Kleene plus over a concatenation of edge labels.
\myQuery\ queries are becoming relevant in real-world applications, such as social networks, financial transactions and SPARQL endpoints.
In order to efficiently process \myIndex\ queries online, we propose the \myIndex\ index, the first reachability index suitable for these queries.
We design an indexing algorithm with pruning rules to not only accelerate the index construction but also remove redundant index entries. 
Our comprehensive experimental study demonstrates that the \myIndex\ index allows to strike a balance between online processing (full online traversals) and offline computation (a materialized transitive closure).
Last but not least, our open-source implementation of the \myIndex\ index can significantly speed up the processing of \myQuery\ queries in current mainstream graph engines.

%% file: sections/Appendix.tex
\appendices

\section{Proof of Theorem \ref{theorem: 2k search}}

A proof of Theorem \ref{theorem: 2k search} is provided in this section.
Before that, we first show the kernel of a label sequence is unique in Lemma \ref{lemma: unique kernel} that will be used in the proof of Theorem \ref{theorem: 2k search}.

\begin{lemma}\label{lemma: unique kernel}
If $L$ has a kernel, then the kernel is unique.
\end{lemma}
\begin{IEEEproof}
The proof is based on induction. 
The statement is if a label sequence $L$ of length $n$ has a kernel, then the kernel is unique.
It is trivial to prove the initial case $|L|=2$.
Assuming the case $n$ is true, then we show the case $n+1$ is also true.
Let $|L|=n+1$. We use $\Bar{L}$ to denote the label sequence obtained by removing the last label of $L$, \textit{i.e.}, $|\Bar{L}|=n$.
We prove the case $n+1$ below.

Assuming $L$ can have two different kernels $L_1$ and $L_2$, and $L_1\neq L_2$, then
 $L=(L_1)^{h_1}\circ L_1',h_1\geq 2$ and $L=(L_2)^{h_2}\circ L_2',h_2\geq 2$.
If $|L'_1|= 0$ and $|L'_2|=0$, then $L$ has two MRs, which is contradictory (Lemma \ref{lemma: unique MR}).
If $|L'_1|\neq 0$ and $|L'_2|\neq 0$, then $\Bar{L}$ has kernels $L_1$ and $L_2$, which contradicts to the case $n$.
The remaining cases are that only one of $L'_1$ and $L'_2$ has a length of $0$. 
W.l.o.g. consider $|L'_1|=0$ and $|L'_2|\neq 0$.
Given this, if $h_1>2$, then $\Bar{L}$ also has kernels $L_1$ and $L_2$, which is contradictory.
Then we have $h_1=2$ and $|L'_1|=0$, \textit{i.e.}, 
\begin{equation}\label{equation: form 1}
L = L_1\circ L_1.     
\end{equation}
In addition, we have 
\begin{equation}\label{equation: form 2}
 L=(L_2)^{h_2}\circ L_2',h_2\geq 2, |L'_2|\neq 0.   
\end{equation}
Let $L=(l_1,...,l_{2|L_1|})$ and $|L_1|=a|L_2|+b,1 \leq a,b < |L_2|$.
According to Equ. (\ref{equation: form 1}), we have $l_i=l_{i+|L_1|},1\leq i \leq |L_1|$ and $l_{i'}=l_{i'-|L_1|},|L_1|< i'\leq 2|L_1|$, which means $l_i=l_{i+a|L_2|+b}$ and $l_{i'}=l_{i'-a|L_2|-b}$.
Based on Equ. (\ref{equation: form 2}), we have $l_i=l_{i+b}$ and $l_{i'}=l_{i'}-b$.
Given this, consider the following two cases: case (i) if $2|L_1|\bmod b =0$, then $|MR(L_1\circ L_1
)|=b\neq |L_1|$ that contradicts to the fact that $L_1$ is the unique MR of $L$;
case (ii) if $2|L_1|\bmod b \neq 0$, then $L=(L_3)^{h_3}\circ L'_3$, where either $|L_3|=b, h_3\geq 2$, and $|L'_3|\neq 0$, or $|L_3|<b$ and $h_3>2$. 
Note that, in the two sub-cases of case (ii), $L'_3$ is  $\epsilon$, or a proper prefix of $L_3$.
Therefore, $\Bar{L}$ has a kernel $L_3$, $|L_3|\leq b$.
However, $\Bar{L}$ also has a kernel $L_2$ and $|L_2|>b\geq |L_3|$, which is also a contradiction.
\end{IEEEproof}

Based on Lemma \ref{lemma: unique kernel}, we prove Theorem \ref{theorem: 2k search} below.


\begin{IEEEproof}[\textbf{Proof of Theorem \ref{theorem: 2k search}}]
It is not difficult to prove Case 1 and Case 2. We focus on Case 3 below.
For ease of presentation, let $\Lambda(u,x)=\Lambda(p(u,x))$ and $\Lambda(x,v)=\Lambda(p(x,v))$.

(Sufficiency) Because $\Lambda(u,x)$ has the kernel $L'$ and the tail $L''$, such that $\Lambda(u,x)=(L')^h\circ L''$, $h\geq 2$. Thus, we have $MR(\Lambda(u,x)\circ \Lambda(x,v))=MR((L')^h\circ L''\circ \Lambda(x,v))= L'$, 
otherwise $MR(L''\circ \Lambda(x,v))\neq L'$.
In addition, we have $|L'|\leq k$ because $|\Lambda(u,x)|=2k$. Thus, $L'$ is the k-MR of $p$.

(Necessity) 
We show that $p$ does not have a non-empty k-MR in the following two cases.
\begin{itemize}[leftmargin = *]
    \item[-] Case (i): $\Lambda(u,x)$ does not have a kernel and a tail.
    Assume that $p$ can have a non-empty k-MR $L'''$ in this case. Because $|L'''|\leq k$ and $|\Lambda(u,x)|=2k$, then $\Lambda(u,x)$ has a kernel and a tail, which is contradiction to the case definition. 

    \item[-] Case (ii): $\Lambda(u,x)$ has a kernel $L'$ and a tail $L''$, but $MR(L''\circ \Lambda(x,v))\neq L'$. Assume that $p$ has a non-empty k-MR $L'''$ in this case. Knowing that $|L'''|\leq k$ and $|\Lambda(u,x)|=2k$, we have $L'''=L'$ because the kernel of $\Lambda(u,x)$ is unique (Lemma \ref{lemma: unique kernel}). Therefore, $MR((L')^h\circ L'' \circ \Lambda(x,v))=L',h\geq 2$, which means $MR(L''\circ \Lambda(x,v))= L'$, that is also a contradiction.
\end{itemize}
\end{IEEEproof}

\section{Proof of Theorem \ref{theorem: condensed myIndex} and \ref{theorem: correctness of myIndex}}
We prove Theorem \ref{theorem: condensed myIndex} and Theorem \ref{theorem: correctness of myIndex} in this section.
Before proceeding further, we first present the following lemmas that will be used to prove the two theorems.

\begin{lemma}\label{lemma: record by reachability of KBS}
Given a path $p(s,t)$ having a k-MR $L$.
If the KBS from $s$ can visit $t$ (or the KBS from $t$ can visit $s$), then the k-MR $L$ of $p(s,t)$ must be recorded in the index.
\end{lemma}
\begin{IEEEproof}
If the KBS from $s$ can visit $t$, then regardless of whether PR1 or PR2 is applied, the k-MR $L$ of $p(s,t)$ must be recorded.
\end{IEEEproof}

\begin{lemma}\label{lemma: path concatenation}
Given two paths $p(s,u)$ and $p(u,t)$ with k-MR $L$ in a graph, where $aid(u) < aid(s)$ and $aid(u) < aid(t)$.
We have: if $aid(u) \leq i$, then the k-MR of the path from $s$ to $t$ through vertex $u$ is recorded by Algorithm \ref{algo: indexing algorithm} in the $i$-th snapshot of the \myIndex\ index that is computed after performing KBS from a vertex with access id $i$.
\end{lemma}
\begin{IEEEproof}
The proof is based on induction. 
It is trivial to prove the initial case $i = 1$.
We assume the case $i= n$ is true and prove the case $i= n+1$ below.
We only need to show the case $aid(u) = n+1$.
Let $p(s,t)=(s,...,u,...,t)$. 

Assuming the backward KBS from $u$ does not visit $s$.
Then PR3 is triggered, such that there exists vertex $w$, $aid(w) < aid(u)$, and  $p(s,w)$ and $p(w,u)$ have the k-MR $L$. This case can be reduced to the case $i=n$, because the k-MR of path $(w,...,u,...,t)$ is $L$ and $aid(w) < aid(u) <n+1$.
In the same way, we can also prove the case if the forward KBS from $u$ does not visit $t$.

We consider the case that both the backward KBS and the forward KBS from $u$ can visit $s$ and $t$.
For $p(s, u)$, we have the following two cases: 
Case (1) $\exists (u,L)\in \lout(s)$; 
Case (2) $\exists (v,L)\in \lout(s)$ and $\exists (v, L)\in \lin(u)$, $aid(v)< aid(u)$. 
For Case (2), both $p(s,v)$ and $p(v,t)$ have the k-MR $L$, such that this case can be reduced to the case $i=n$ as $aid(v) < aid(u) = n+1$. 
Then we only need to consider Case (1), \textit{i.e.}, $\exists (u,L)\in \lout(s)$.
In the same way, for $p(u, t)$, we only need to consider the case $\exists (u, L)\in \lin(s)$.
Given this, the k-MR of the path from $s$ to $t$ is recorded by having $ (u,L)\in \lout(s)$ and $ (u, L)\in \lin(t)$.
\end{IEEEproof}

\begin{lemma}\label{lemma: PR3}
Given a path $p$ from $s$ to $t$ with a k-MR $L$. 
If the index entry $(t,L)\in \lout(s)$ (or $(s,L)\in \lin(t)$) is pruned because of PR3, then we have one of the following two cases:
\begin{itemize}[leftmargin = *]
    \item[-] $\exists (s, L)\in \lin(t)$ (or $\exists (t, L)\in \lout(s)$); 
    \item[-] $\exists (v,L)\in \lout(s)$ and $\exists (v,L) \in \lin(t)$, such that $aid(v) < aid(t)$ (or $aid(v) < aid(s)$).
\end{itemize}
\end{lemma}
\begin{IEEEproof}
There exists two cases: $aid(t)\leq aid(s)$ or $aid(t)> aid(s)$.
We prove the case of $aid(t)\leq aid(s)$. The proof for the other case follows the same sketch.
Let $p(s,t)=(s,...,u,...,t)$, such that PR3 can be triggered. W.l.o.g. let $aid(u) < aid(t)$ (if $aid(u) \geq aid(t)$ and PR3 is triggered, then there exists vertex $w,aid(w)<aid(u)$, which can be reduced to the case of $aid(u) < aid(t)$).
Given this, we have path $p(s,u)$ and $p(u,t)$ have the same k-MR $L$ according to the definition of PR3.
Then we have three cases: Case (1) $aid(s) > aid(u)$; Case (2) $aid(s) = aid(u)$; Case (3) $aid(s) < aid(u)$.
Case (1) can be proved by Lemma \ref{lemma: path concatenation}, because $aid(s) > aid(u)$, $aid(t) > aid(u)$, and both $p(s,u)$ and $p(u,t)$ have k-MR $L$.
Case (2) can be proved by Lemma \ref{lemma: record by reachability of KBS} because the backward KBS from $t$ can visit $u$, \textit{i.e.}, $aid(s) = aid(u)$.
Case (3) can also by proved by \ref{lemma: record by reachability of KBS} if the forward KBS from $s$ can visit $t$. 
The only case left is that $aid(s) < aid(u)$ and the forward KBS from $s$ cannot visit $t$ because of PR3.
In this case, there must exist vertex $v, aid(v) < aid(s) < aid(u) = n+1$,
and the k-MR of $p(s,v)$ and $p(v,u)$ is $L$.
Then we have $aid(v) < aid(s)$ and $aid(v) <aid(t)$, and both path $p(s,v)$ and $(v,...,u,...,t)$ have the k-MR $L$, which can be proved by Lemma \ref{lemma: path concatenation}.
\end{IEEEproof}

\begin{IEEEproof}[\textbf{Proof of Theorem \ref{theorem: condensed myIndex}}]
Assuming there exists index entry $(t,L)\in \lout(s)$ in the \myIndex\ index , and there also exist $(u,L)\in \lout(s)$ and $(u,L)\in \lin(t)$.
Then we have $aid(u) \geq aid(t)$, otherwise $(t,L)\in \lout(s)$ can be pruned.
Given this, $(u,L)\in \lin(t)$ cannot exist because the backward KBS from $t$ is performed earlier than the forward KBS from $u$, which means we have either $(t,L)\in \lout(u)$, or $(v,L)\in \lout(u)$ and $ (v,L)\in \lin(t)$, such that $(u,L)\in \lin(t)$ is pruned.
The proof follows the same sketch if $(s,L)\in \lin(t)$ is considered.
\end{IEEEproof}


\begin{IEEEproof}[\textbf{Proof of Theorem \ref{theorem: correctness of myIndex}}]
    (Sufficiency)
    It is straightforward.
    
    (Necessity)
    Let $p$ be the path from $s$ to $t$ with the k-MR $L$. W.l.o.g. let the backward KBS from $t$ be performed first.
    Then we have two cases: the backward KBS from $t$ can visit or cannot visit $s$.
    In the first case, the k-MR $L$ of path $p$ must be recorded according to Lemma \ref{lemma: record by reachability of KBS}.
    In the second case, PR3 must be triggered. According to Lemma \ref{lemma: PR3}, we have the k-MR of $p$ is also recorded. 
\end{IEEEproof}

\begin{figure}
    \centering
    \includegraphics[width=\linewidth]{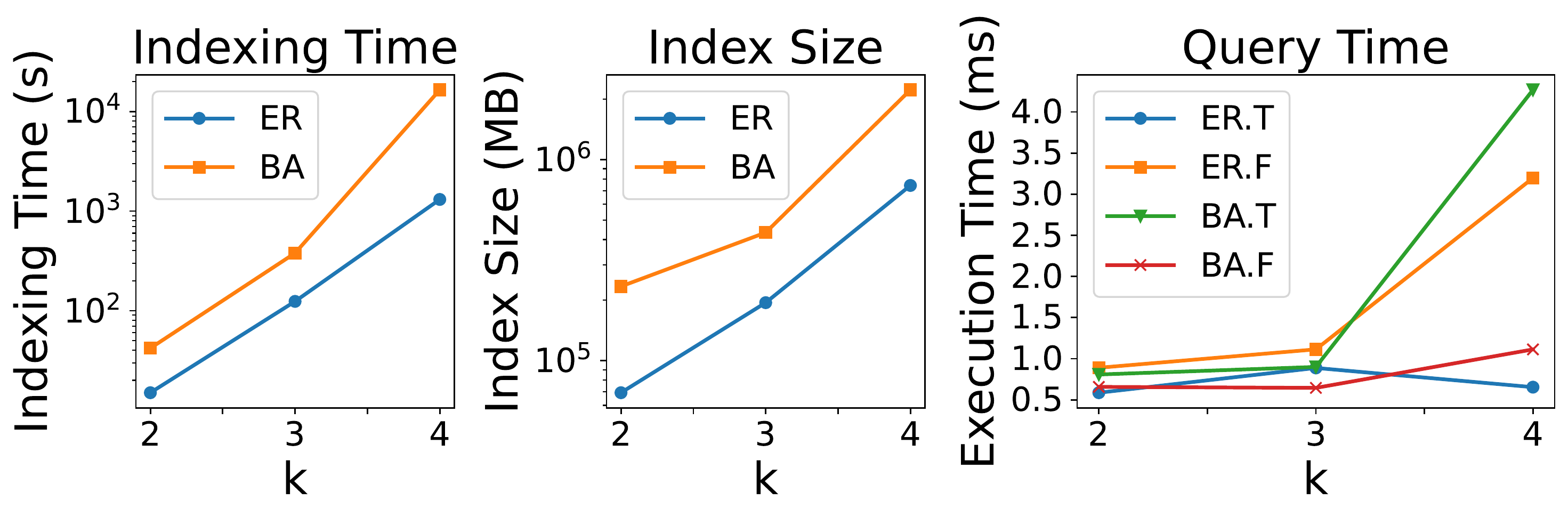}
    \caption{Evaluation of the \myIndex\ index with varying $k$.
    }
    \label{fig: various k}
\end{figure}

\section{Impact of k}\label{subsubsec: impact of k}
In this experiment, we aim at analyzing the impact of $k$ on the index. 
We use a BA-graph and an ER-graph with 125K vertices, average degree $5$, $16$ edge labels. We build \myIndex\ indexes for the BA-graph and the ER-graph with $k$ in $(2,3,4)$.
For each graph, we evaluate a true-query set of 1000 queries and a false-query set of 1000 queries using the three different indexes built with the three different $k$ values.
The results of indexing time, index size, and query time are reported in Fig. \ref{fig: various k}.

\textit{Overall results.}
Fig. \ref{fig: various k} shows that the indexing time and index size of both types of synthetic graphs rise exponentially as $k$ grows. 
The fundamental reason is that as $k$ increases, the number of possible minimum repeats that have to be considered in graph traversals during indexing increases exponentially, which is an inherent step in building a complete reachability index for \myQuery\ queries.
The exponential increase of minimum repeats w.r.t $k$ also has an impact on query time, particularly on the true-queries of BA-graphs and the false-queries on ER-graphs.
This is mainly due to larger index size, and the true-queries on BA-graphs and the false-queries on ER-graphs are more expensive to process than their opposites, respectively.

\section{Remarks}
An alternative version of the \myIndex\ index allowed to concatenate different minimum repeats to answer an \myQuery\ query, \textit{i.e.}, in Case 1 of Definition \ref{definition: myIndex}, $L'$ can be different from $L''$ in the alternative version. 
However, such a design will prevent the use of PR3 that can prune vertices and avoid redundant traversals.
The reason is related to the technical proof of Theorem 3. In a nutshell, if concatenating different minimum repeats is allowed and PR3 is also allowed to apply, then the index might not be complete, i.e., missing some index entries. Therefore, we can only allow one of the two designs, \textit{i.e.}, either allowing concatenating different minimum repeats or allowing applying PR3. In the previous version, we allowed the former one.
Consequently, the indexing time of the alternative version is much longer than the version introduced in this paper. For instance, for the smallest graph used in our experiments (the AD graph presented in Section \ref{section: experimental results}), the indexing time of the alternative version is 32x slower than the current one. Thus, we focus on concatenating the same minimum repeats.